\documentclass[11pt]{article}
\usepackage{jheppub}
\usepackage{hyperref}
\usepackage{amsmath,amssymb,amsthm,amscd}
\usepackage[all,cmtip]{xy}
\usepackage{xcolor}
\usepackage{blkarray}
\usepackage{float} 
\usepackage{youngtab}
\usepackage{mathtools}
\usepackage{comment}
\usepackage{braket}
\usepackage{ytableau}
\usepackage{subcaption}
\usepackage[normalem]{ulem}
\Ylinethick{0.3pt}
\Yboxdim{5pt}
\usepackage{tikz,textcomp}
\usetikzlibrary{decorations.pathreplacing,calc,fadings,fit,shapes,arrows,positioning,chains,matrix}

\DeclareFontFamily{U}{wncy}{}
\DeclareFontShape{U}{wncy}{m}{n}{<->wncyr10}{}
\DeclareSymbolFont{mcy}{U}{wncy}{m}{n}
\DeclareMathSymbol{\Sh}{\mathord}{mcy}{"58} 

\numberwithin{equation}{section}

\def\bea{\begin{eqnarray}}
\def\eea{\end{eqnarray}}
\def\be{\begin{equation}}
\def\ee{\end{equation}}
\def\ba{\begin{align}}
\def\ea{\end{align}}
\def\bse{\begin{subequations}}
\def\ese{\end{subequations}}
\newcommand{\beq}{\begin{equation}\begin{aligned}}
\newcommand{\eeq}{\end{aligned}\end{equation}}

\newcommand{\sech}{\rm sech}
\newcommand{\RR}{\mathbb{R}}

\definecolor{darkgreen}{rgb}{0.0, 0.5, 0.2}

\title{Confinement Versus Screening in the Schwinger Model on AdS$_2$ from Bosonization and Tensor Networks}
\vspace{2em}
\abstract{
We analyze confinement and screening in single-flavor quantum electrodynamics (QED$_2$) on two-dimensional anti-de Sitter space (AdS$_2$), with and without a Schwarzschild black hole, both in the continuum and on the lattice. The theory is formulated in two frames adapted to distinct choices of a preferred time coordinate: the Schwarzschild frame, associated with the Boulware vacuum, and the global AdS$_2$ frame, associated with the $\mathrm{SL}(2,\mathbb{R})$-invariant vacuum. In the massless limit, the static potential between an external charge-anticharge pair is obtained in closed form by bosonization, at both zero and finite temperature. After subtraction of the position-dependent probe self-energies, which, unlike in flat space, are not constant, the potential remains finite as the geodesic separation is taken to infinity, establishing that the theory is screened. This is consistent with the explicit breaking of the $\mathrm{U}(1)$ electric one-form symmetry by the dynamical fermions, and resolves a confining/screening ambiguity in earlier treatments that identify the static potential with the unsubtracted ground-state energy.  To validate the continuum analysis, we propose a covariant discretization scheme for placing fermions in curved spacetime on the lattice while ensuring that the continuum properties of the spin and gauge connections are restored in the continuum limit. This construction resolves ambiguities in the existing literature on lattice fermions in curved backgrounds and provides the foundation for our tensor-network simulations. Using a matrix product state ansatz, we confirm our analytical predictions for the phase diagram in AdS$_2$. We perform extensive numerical simulations of the static potential and the electric flux-tube profile for varying fermion masses, which we match to the continuum prediction.
}

\author[a,b]{Sriram Bharadwaj}
\author[a,b]{, Jack Isen}
\author[a,b,c]{, Zhong-Bo Kang}

\affiliation[a]{Department of Physics and Astronomy, University of California, Los Angeles, CA 90095, USA}
\affiliation[b]{Mani L. Bhaumik Institute for Theoretical Physics, University of California, Los Angeles, CA 90095, USA}
\affiliation[c]{Center for Quantum Science and Engineering, University of California, Los Angeles, CA 90095, USA}

\emailAdd{sbharadwaj@physics.ucla.edu}
\emailAdd{jackisen@physics.ucla.edu}
\emailAdd{zkang@physics.ucla.edu}

\begin{document}

\maketitle

\newpage
\section{Introduction}
Confinement remains one of the central open problems in quantum field theory~\cite{Shifman:2012zz,Greensite2020Confinement}. In $(3+1)$-dimensional non-abelian gauge theories, the flux between two static color charges is squeezed into a tube with a linearly rising potential, so that isolated color charges are never observed; this picture has strong numerical and experimental support but no analytic proof, and a first-principles understanding of the confining vacuum is still lacking~\cite{Greensite2020Confinement}. 

Lattice gauge theory provides a nonperturbative framework for investigating confinement and the dynamics of strongly coupled gauge theories. In the seminal work~\cite{Wilson1974}, Wilson developed the Euclidean formulation of lattice gauge theories, which is suited to Monte Carlo simulations of observables such as static interquark potentials and Polyakov loops~\cite{Creutz1980,Polyakov1978}. Conventional Euclidean Monte Carlo methods, however, face limitations for real-time dynamics and systems at finite fermion density, where the path-integral weights become oscillatory or complex~\cite{TroyerWiese2005,deForcrand2009}. The Kogut-Susskind Hamiltonian formulation~\cite{KogutSusskind1975} provides an alternative framework that is suited to Hamiltonian simulation~\cite{ZoharCiracReznik2016,BanulsEtAl2020}. Quantum simulation has emerged as a powerful tool to study quantum field theories in the Hamiltonian formulation due to its ability to circumvent the sign problem that plagues Monte Carlo methods~\cite{TroyerWiese2005,JordanLeePreskill2012,BanulsEtAl2020}. Moreover, it enables the experimental realization of lattice gauge theories on near-term quantum hardware~\cite{MartinezEtAl2016,KlcoEtAl2018,GonzalezCuadra:2024StringBreaking, John:2026NonAbelianStringBreaking}. This is significant in two and three spatial dimensions, where classical computation struggles due to the exponential growth of the Hilbert-space dimension~\cite{Feynman1982,BanulsEtAl2020}.

The situation is far more tractable in $(1+1)$ dimensions both analytically and numerically, where Gauss's law can be solved explicitly and the gauge field has no local propagating degrees of freedom. The paradigmatic example is the Schwinger model, QED$_2$ with a single fermion flavor~\cite{Schwinger:1962tp}, whose massless limit is exactly solvable by bosonization. This model has served as a canonical laboratory for nonperturbative phenomena including anomaly-induced mass generation, vacuum polarization, $\theta$-vacua, charge screening, and confinement~\cite{Schwinger:1962tp,LowensteinSwieca1971,Coleman:1975pw,Coleman:1976uz}.  On the simulation side, the $(1+1)$-dimensional setting is particularly well-suited to tensor-network methods, which provide an efficient description of low-energy states and enable controlled simulations of spectra, thermal observables, and real-time dynamics~\cite{Schollwock2011,ByrnesEtAl2002,BanulsEtAl2013,BuyensEtAl2014}. This combination of analytical control and efficient Hamiltonian simulation makes QED$_2$ a natural laboratory for asking how confinement and screening are modified on curved backgrounds.

A central feature of quantum field theory in curved spacetime is that many statements which are unambiguous in flat space become dependent on a preferred choice of time-like Killing vector and, hence, the associated vacuum~\cite{Spradlin:1999bt}. The presence of Hawking~\cite{Hawking:1975vcx} and Unruh~\cite{Unruh:1976db} radiation, the entanglement entropy~\cite{FuentesSchuller:2004xp}, and the very notion of a particle~\cite{Fulling:1972md} are all dependent on this choice. For the AdS$_2$ black hole of interest here, the Boulware, Hartle-Hawking, and $\mathrm{SL}(2,\mathbb{R})$-invariant vacua and their relations were analyzed in~\cite{Spradlin:1999bt}. 

Given the need to specify a preferred time, it is important to analyze phenomena in quantum field theory that are sensitive to this choice. In particular, we ask whether confinement in QED$_2$, as diagnosed through the static potential, is dependent on the choice of preferred time in AdS$_2$; this problem has a subtle history~\cite{CallanWilczek1990}. It has been suggested~\cite{Alimohammadi:2000fg, MohseniSadjadi:2000nt, MohseniSadjadi:2000mg} that the Schwinger Model in two-dimensional curved space could be confining or screening, depending on precisely how the geodesic separation (i.e. the proper distance between two probe particles at fixed coordinate time) is taken to infinity, in apparent tension with the explicitly broken electric one-form symmetry. We resolve these ambiguities in QED$_2$ on AdS$_2$ by computing the vacuum energy, static potential, and flux-tube profile in the presence of probe charges, carefully accounting for self-energy contributions of the probe charges. 

\subsection{Main Results}
First, we clarify precisely what we mean by ``screened" and ``confined", which we define following~\cite{Gross:1995bp}. A theory is confined if the binding energy between a pair of equal and opposite probe charges diverges with their geodesic separation. On the other hand, a theory is screened if the probe-pair binding energy instead stays finite as their geodesic separation diverges.

We analyze whether single-flavor QED$_2$ on AdS$_2$, with and without a Schwarzschild black hole, screens or confines in the continuum and on the lattice. Our primary analytical tool is bosonization: the massless Schwinger model maps to a massive dual scalar. Obtaining a semiclassical solution to the Hamiltonian's equations of motion in the presence of probe charges yields the static potential and the electric flux-tube profile in closed form. On the lattice, we formulate the theory as a qubit Hamiltonian and perform a Hamiltonian simulation using tensor networks (a matrix product state ansatz optimized by the Density Matrix Renormalization Group (DMRG) algorithm), which is highly efficient for gapped $(1+1)$D systems~\cite{White:1992zz,Schollwock:2010zz} and gives us direct access to the ground state in the presence of probe charges. Throughout, we work in two frames adapted to distinct choices of preferred time: the Schwarzschild frame, as in \eqref{eq:SchwarzchildBH}, and the global AdS$_2$ frame, as in \eqref{eq:globalAdS}. Our key contributions are as follows.

\begin{enumerate}
    \item \textbf{Covariant discretization for lattice gauge theories with dynamical fermions in curved space.} We develop a systematic covariant discretization scheme by packaging the spin and gauge connections together with the kinetic term into a single generalized covariant derivative \eqref{eq:CovDer}, which we then discretize. In a curved background, the kinetic and spin-connection terms are each non-Hermitian and become Hermitian only in combination; discretizing them separately spoils this cancellation at finite lattice spacing. Our construction is manifestly Hermitian at finite spacing and reproduces the correct continuum Hamiltonian. This resolves existing ambiguities in the lattice-fermion literature \cite{ikeda2025quantumsimulationfermionsads2, ikeda2026geometryinducedchiraltransport} and fills a long-standing gap~(Sec.~\ref{sec:Hamiltonian}).

    \item \textbf{Continuum resolution of the screening-versus-confinement puzzle.} The phase of QED$_2$ on AdS$_2$ has a confusing history: the massless theory has been claimed to confine~\cite{Alimohammadi:2000fg, MohseniSadjadi:2000mg, MohseniSadjadi:2000nt, Gass:AdS2, Ghosh:1996}, in apparent tension with the fact that the dynamical fermions explicitly break the $\mathrm{U}(1)$ electric one-form symmetry of Maxwell theory, which is expected to screen any integer-charge flux tube. We settle this by solving the bosonized semiclassical theory exactly for the vacuum energy in the presence of probe charges. In contrast to flat space, the crucial point is that the probe self-energies are no longer translation-invariant constants. Once their position-dependent values are carefully subtracted, the binding potential is screened, in harmony with the absence of a $\rm U(1)$ global one-form symmetry. We carry out this calculation in both the Schwarzschild and global AdS$_2$ frames, where the previous literature reached seemingly contradictory conclusions, and we explain why the two descriptions are in fact compatible. We obtain closed-form expressions for both the static $q\bar q$ potential and the electric flux-tube profile, at zero and finite temperature~(Sec.~\ref{sec:Screening}).

    \item \textbf{The screened phase from Hamiltonian simulations with tensor networks.} The covariant discretization is applied to numerically validate the predicted screened phase. We compute the static binding potential on the lattice and, after careful numerical subtraction of the position-dependent self-energy contributions, we confirm that the theory is screened at finite lattice spacing, in agreement with the continuum expectation. Furthermore, we extract the electric flux-tube profile and find strong agreement with the continuum bosonization result. We also obtain results for the massive theory, which is analytically tractable only in the small-mass expansion \cite{Alimohammadi:2000fg}. We conclude with a real-time simulation of a string breaking process starting from an initial state with a dynamical electron-positron pair connected by a Wilson line. This serves as a non-perturbative check of the screened phase (Sec.~\ref{sec:Simulation}).
\end{enumerate}

\subsection{Relation to prior works and novelty}
Our results should be contrasted with the existing literature~\cite{Boada:2010sh,Villegas:2014dqa,Yang:2019kbb,Lewis:2019oyx} on the massless Schwinger model in curved space, both in the continuuum and on the lattice. 

In the continuum, the static potential in pure AdS$_2$ and dS$_2$ has been studied \cite{Alimohammadi:2000fg, MohseniSadjadi:2000mg, MohseniSadjadi:2000nt, Gass:AdS2}, and for generic backgrounds specialized to the Schwarzschild case \cite{Ghosh:1996}. These works identify the static potential directly with the ground-state energy in the presence of the probe charges, and on this basis reach conclusions that depend on how the charges are taken to infinity, and that appear, in places, mutually inconsistent. In curved space, however, this identification requires some care: unlike in flat space, the probe self-energies are not translation-invariant constants, and must be subtracted before the binding potential can be read off. 
After this subtraction, the different limits give a single screened phase, in agreement with the one-form symmetry argument, and the apparent tension between them is resolved.

On the lattice, our construction clarifies a subtle point in the staggered-fermion Hamiltonian of~\cite{ikeda2025quantumsimulationfermionsads2}. In a curved background, the kinetic term and the spin-connection term are not separately Hermitian, but become Hermitian only in combination, as also noted in that work. This makes it important to discretize the combined covariant derivative directly. We implement this by packaging the spin and gauge connections into a single Hermitian covariant derivative before discretization (see Sec.~\ref{sec:Hamiltonian}), leading to a lattice Hamiltonian that is manifestly Hermitian at finite lattice spacing and reproduces the correct continuum limit.

\subsection{Outline}
The remainder of the paper is organized as follows. In Sec.~\ref{sec:Hamiltonian} we derive the gauge-fixed QED$_2$ Hamiltonian on AdS$_2$-Schwarzschild in both the Schwarzschild and global AdS$_2$ frames, passing from the continuum to the staggered lattice and finally to a qubit model, and discuss the fate of chiral symmetry. In Sec.~\ref{sec:Screening} we compute the static potential and Wilson loop analytically and establish the screened phase of the massless theory at both zero and finite temperature. In Sec.~\ref{sec:Simulation} we validate the screened phase of QED$_2$ on AdS$_2$ with DMRG simulations of the static potential, the electric flux-tube profile, and real-time string breaking processes. Appendix~\ref{app:GaugeInvariance} establishes the diffeomorphism (gravitational gauge) invariance of the Hamiltonian in the continuum and on the lattice.

\section{QED$_2$ in an AdS$_2$-Schwarzschild background}\label{sec:Hamiltonian}
In this section, we formulate the gauge-fixed Hamiltonian of $(1+1)$D Quantum Electrodynamics (QED$_2$) in the presence of a black-hole background in negatively curved AdS$_2$ spacetime. The continuum Hamiltonian is then discretized via staggered fermions after integrating out the Gauss constraint. The lattice Hamiltonian is then mapped to a system of qubits via a Jordan-Wigner transformation, which serves as the starting point for our numerical studies in Section~\ref{sec:Simulation}. We carry out the discretization with respect to two distinct frames, the Schwarzschild frame and the global AdS$_2$ frame. 

\subsection{Spacetime geometry}
To analyze the problem of confinement in the Hamiltonian formulation, it is essential to consider the static potential between a pair of oppositely charged probe particles. The notion of a ``static" charge is subtle in curved space. In flat space, when one considers a pair of static charges, they are typically also taken to have zero (proper) acceleration. However, this is generically not the case in curved spacetime. 

\paragraph{Pure AdS$_2$ in Schwarzschild coordinates.} 
We define AdS$_2$ in Schwarzschild coordinates
\begin{align}\label{eq:SchwarzchildBH}
    ds^2 &= -f(r)dt^2 + \frac{dr^2}{f(r)}, &f(r)= \frac{r^2}{L^2}.
\end{align}
Note that the metric in equation \eqref{eq:SchwarzchildBH} has a coordinate singularity at $r=0$. We now analyze the proper acceleration in the Schwarzschild frame at a constant radial coordinate $r$. For such a trajectory, the spatial velocity vanishes ($u^r = 0$), and the velocity is normalized by $u^\mu u_\mu = -1$, yielding $u^T = 1/\sqrt{f(r)}$. The proper acceleration vector is given by the covariant derivative of the velocity along the particle's worldline, $a^\mu = u^\nu \nabla_\nu u^\mu$. Because the observer is stationary and the metric components are independent of $T$, this reduces to $a^\mu = \Gamma^\mu_{TT} (u^T)^2 = \Gamma^\mu_{TT} / f(r)$. Calculating the invariant scalar magnitude $a = \sqrt{a_\mu a^\mu}$ for this metric yields
\begin{equation}
    a(r) = \frac{f'(r)}{2\sqrt{f(r)}} = \frac1L.
\end{equation}
Instead, if one considers the black hole background with $f(r) = (r^2 - r_h^2)/L^2$, the scalar proper acceleration is strictly positive everywhere outside the horizon ($r > r_h$) and diverges as $r \to r_h$. This non-zero acceleration means a particle must be constantly acted on by a force to counteract gravity and maintain a fixed radial position.

\paragraph{The global AdS$_2$ frame.} The metric in the global AdS$_2$ frame is:
\begin{equation}\label{eq:globalAdS}
    ds^2 = L^2 \left( -\cosh^2\rho \, d\tau^2 + d\rho^2 \right).
\end{equation}
Consider an observer at the spatial origin of the global space, $\rho = 0$. Their 4-velocity is $u^\mu = (1/L, 0)$. The relevant Christoffel symbol for their acceleration is $\Gamma^\rho_{\tau\tau} = \sinh\rho \cosh\rho$. Evaluated at $\rho = 0$, the Christoffel symbol vanishes exactly, yielding $a^\mu = 0$. 

\subsection{The action and the vacuum} 

To proceed with our analysis of confinement in curved space, we need a covariant starting point, which is given by the gauge and diffeomorphism invariant action. We first introduce the curved and tangent space gamma-matrices, which allows us to introduce a spin connection that is used to covariantly put the fermionic theory on a curved manifold. Following this, we write the QED$_2$ action in curved space, which is manifestly $\rm U(1)$ and diffeomorphism invariant. To complete this analysis, following~\cite{Spradlin:1999bt}, we describe how the AdS$_2$ background geometry arises as the near-horizon limit of $(3+1)$-dimensional Reissner-Nordstrom black hole, which determines the vacuum of the theory.

To couple the fermionic fields to the curved background, we introduce the zweibein $e^a_\mu$, in terms of which one defines the curved-space $\Gamma$-matrices:
\begin{align}
    \{\Gamma^{\mu}, \Gamma^{\nu}\} &= -2g^{\mu\nu}, &\Gamma^\mu& = e^{\mu}_a \gamma^a, &e^{a}_\mu e^{\mu}_b &= \delta_{b}^a.
\end{align}
We adopt the tangent-space representations $\gamma^0= \sigma^z$, $\gamma^1 = i\sigma^y$, and $\gamma^5 = \gamma^0\gamma^1 = \sigma^x$, which satisfy:
\begin{align}
    \{\gamma^a, \gamma^b\} = -2\eta^{ab}.
\end{align}
To go between the tangent-space indices ($a, b, \dots = 0, 1$) and spacetime  indices ($\mu, \nu, \dots = t, r$ for Schwarzschild and $\mu, \nu, \dots = \tau,\rho$ for global AdS$_2$), we use the zweibein $e^a_{\mu}$ and $e_{a}^{\mu}$ respectively. Explicitly, 
\begin{align}
    g^{\mu\nu} = e^{\mu}_a \,e^{\nu}_b\,\eta^{ab}.
\end{align}

\paragraph{QED$_2$ action.} The Dirac covariant derivative relies on the spin connection $\omega^{ab}_\mu$, which is completely determined by imposing the torsion-free Maurer-Cartan equation:
\begin{align}
    de^a+\omega^{ab}\wedge e_b=0.
\end{align}
The fermionic action is
\begin{align}
    S_f[A,\psi,\bar\psi]
    &=
    \int d^2x\sqrt{-g}\;
    \bar\psi\left(i\Gamma^\mu D_\mu-m\right)\psi, & D_\mu
    &=
    \partial_\mu+iA_\mu-\frac14\omega_\mu^{ab}\gamma_a\gamma_b,
\end{align}
where $\bar \psi\equiv \psi^\dagger \gamma^0$ is defined with the tangent-space $\gamma$-matrices. The gauge action is
\begin{align}
    S_g[A]
    =
    -\frac{1}{4e^2}\int d^2x\sqrt{-g}\;F_{\mu\nu}F^{\mu\nu}.
\end{align}

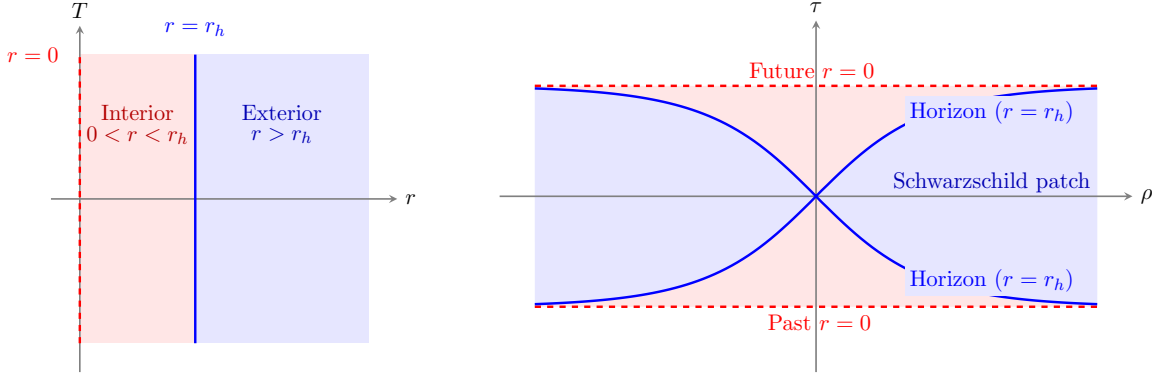
\begin{figure}[h]
    \centering
    \raisebox{-0.5\height}{\resizebox{!}{5cm}{
    \begin{tikzpicture}[>=stealth, scale=1.]
        \fill[red!10] (0, -2.5) rectangle (2, 2.5);
        \fill[blue!10] (2, -2.5) rectangle (5, 2.5);
        \draw[->, thick, gray] (-0.5, 0) -- (5.5, 0) node[right, black] {$r$};
        \draw[->, thick, gray] (0, -3) -- (0, 3) node[above, black] {$T$};
        \draw[red, very thick, dashed] (0, -2.5) -- (0, 2.5) node[above=-10pt, right=-40pt, align=left] {$r=0$};
        \draw[blue, very thick] (2, -2.5) -- (2, 2.5) node[above=5pt, align=center] {$r=r_h$};
        \node[red!70!black] at (1, 1.5) {Interior};
        \node[red!70!black] at (1, 1.1) {$0 < r < r_h$};
        \node[blue!70!black] at (3.5, 1.5) {Exterior};
        \node[blue!70!black] at (3.5, 1.1) {$r > r_h$};
    \end{tikzpicture}
    }}
    \hfill
    \raisebox{-0.5\height}{\resizebox{!}{5cm}{
    \begin{tikzpicture}[>=stealth, scale=1.2]
        \def\halfpi{1.5708}
        \fill[red!10] (-4, \halfpi) -- plot[domain=-4:4, samples=101] (\x, {acos(1/cosh(\x))*3.14159/180}) -- (4, \halfpi) -- cycle;
        \fill[red!10] (-4, -\halfpi) -- plot[domain=-4:4, samples=101] (\x, {-acos(1/cosh(\x))*3.14159/180}) -- (4, -\halfpi) -- cycle;
        \fill[blue!10] (4, 0) -- plot[domain=4:0, samples=101] (\x, {acos(1/cosh(\x))*3.14159/180}) -- plot[domain=0:4, samples=101] (\x, {-acos(1/cosh(\x))*3.14159/180}) -- cycle;
        \fill[blue!10] (-4, 0) -- plot[domain=-4:0, samples=101] (\x, {acos(1/cosh(\x))*3.14159/180}) -- plot[domain=0:-4, samples=101] (\x, {-acos(1/cosh(\x))*3.14159/180}) -- cycle;
        \draw[->, thick, gray] (-4.5, 0) -- (4.5, 0) node[right, black] {$\rho$};
        \draw[->, thick, gray] (0, -2.5) -- (0, 2.5) node[above, black] {$\tau$};
        \draw[red, very thick, dashed] (-4, \halfpi) -- (4, \halfpi);
        \draw[red, very thick, dashed] (-4, -\halfpi) -- (4, -\halfpi);
        \node[red, above] at (-0.06, \halfpi) {Future $r=0$};
        \node[red, below] at (0.05, -\halfpi) {Past $r=0$};
        \draw[blue, very thick, domain=-4:4, samples=101] plot (\x, {acos(1/cosh(\x))*3.14159/180});
        \draw[blue, very thick, domain=-4:4, samples=101] plot (\x, {-acos(1/cosh(\x))*3.14159/180});
        \node[blue, fill=blue!10, inner sep=2pt] at (2.5, 1.2) {Horizon ($r=r_h$)};
        \node[blue, fill=blue!10, inner sep=2pt] at (2.5, -1.2) {Horizon ($r=r_h$)};
        \node[blue!70!black] at (2.5, 0.2) {Schwarzschild patch};
    \end{tikzpicture}
    }}
    \caption{Two coordinate representations of the same spacetime. {(Left):} The regularized static patch coordinates $(T, r)$: in this local frame, the singularity at $r=0$ appears as a vertical spatial boundary, and the bifurcate Killing horizon at $r=r_h$ is represented as a stationary radial location. {(Right):} The same spacetime mapped into global $\text{AdS}_2$ coordinates $(\tau, \rho)$: in this global covering space, the horizon unfolds into intersecting light-like rays ($\cos\tau = \text{sech}\,\rho$), and the $r=0$ point maps directly to spacelike boundaries at constant global time ($\tau = \pm \pi/2$).}
    \label{fig:regularized_coords}
    \label{fig:global_coords}
\end{figure}
\paragraph{The vacuum.} As pointed out in~\cite{Spradlin:1999bt}, an AdS$_2$ black-hole spacetime is an AdS$_2$ spacetime, which is $\rm SL(2, \RR)$-invariant, equipped with a preferred choice of time coordinate. A typical construction\footnote{An equivalent construction involves coupling the gravitational theory to a linear dilaton with a specified profile $\Phi$, that picks out a preferred timelike Killing vector $\xi_\mu = \epsilon_{\mu\nu}\nabla^\nu\Phi$.} involves taking the near-horizon limit of a Reissner-Nordstrom black hole, which yields the AdS$_2$-Schwarzschild case with metric factor $f(r) = (r^2-r_h^2)/L^2$:
\begin{align}\label{eq:SchwarzschildBH-2}
    ds^2 = -\frac{r^2-r_h^2}{L^2}dt^2 + \frac{L^2}{r^2 - r_h^2}dr^2.
\end{align}
The choice of time coordinate $t$ (called the Schwarzschild time) is fixed by the choice of time in the $(3+1)$-dimensional parent theory. 

In the pure AdS$_2$ limit $r_h\rightarrow0$, the metric reduces to the Poincar\'e patch upon the change of variables $r = 1/z$.\footnote{We will refer to pure AdS$_2$ with $r_h=0$ in Schwarzschild coordinates $(t, r)$ as ``Schwarzschild AdS$_2$".} Moreover, the dependence of the metric on the Hawking temperature $T_H = \frac{r_h}{2\pi L^2}$ may be removed by a coordinate transformation, i.e.
\begin{align}
    r &= r_h \cosh\rho \cos\tau, & t &= \frac{L^2}{r_h} \text{arctanh} \left( \frac{\sin\tau}{\tanh\rho} \right).
\end{align}
This yields the so-called global AdS$_2$ metric:
\begin{align}
    ds^2 = L^2 \left( -\cosh^2\rho \, d\tau^2 + d\rho^2 \right).
\end{align}
Interestingly, the vacuum of the quantum theory satisfies~\cite{Spradlin:1999bt}:
\begin{align}
    \ket{\Omega_{\rm Global}} & = \ket{\Omega_{\text{Poincar\'e}}} = \ket{\Omega_{\rm Hartle-Hawking}}, &  \ket{\Omega_{\text{Schwarzschild}}} & = \ket{\Omega_{\text{Boulware}}}.
\end{align}

Although the vacua for the Schwarzschild and global AdS$_2$ patches are equivalent~\cite{Spradlin:1999bt}, their preferred choices of time coordinates, i.e. $t$ versus $\tau$, are distinct. Consequently, the Hamiltonians and energies in these two frames are expected to be different, which is the root of the confinement-versus-screening puzzle in the two frames. In our case, the choice of preferred time will follow the time slice on which we place a pair of probe charges, whose static potential we evaluate.

\subsection{Pure AdS$_2$ Hamiltonian in Schwarzschild coordinates: continuum and lattice}
We now derive the QED$_2$ Hamiltonian in Schwarzschild coordinates. Starting from the continuum Lagrangian, the continuum Hamiltonian is obtained via a Legendre transformation. Then we discretize this Hamiltonian with Kogut-Susskind staggered fermions. Finally, we map to a qubit spin-chain Hamiltonian via  Jordan-Wigner transformation in a form suited to Hamiltonian simulations. 

\paragraph{The Continuum Hamiltonian.}
We begin by writing the fermionic Lagrangian. Because the metric determinant is constant ($\sqrt{-g}=1$), the integration measure is trivial. We work in the temporal gauge $A_t=0$. The covariant derivatives then take the form:
\begin{align}
    D_t &= \partial_t+\frac14 f'(r)\gamma^5, & D_r &= \partial_r+iA_r.
\end{align}
Using the curved-space Dirac matrices derived previously, we evaluate the kinetic operators:
\begin{align}
    \gamma^0\Gamma^t &= \frac{1}{\sqrt{f(r)}}, \qquad \gamma^0\Gamma^r = \sqrt{f(r)}\,\sigma^x.
\end{align}
This allows us to explicitly expand the fermionic Lagrangian density:
\begin{align}
    \mathcal L_f = \psi^\dagger \left( \frac{i}{\sqrt{f(r)}}\partial_t +i \sqrt{f(r)}\,\sigma^x\left(\partial_r+iA_r\right) +i\frac{f'(r)}{4\sqrt{f(r)}}\sigma^x -m\sigma^z \right)\psi.
\end{align}
Rescaling the fermionic field by the spatial factor, $\chi=f(r)^{-\frac14}\psi$ transforms the spatial derivative via the product rule:
\begin{align}
    \left(\partial_r-iA_r\right)\psi = f(r)^{\frac14}\left(\partial_r+iA_r\right)\chi +\frac14 f(r)^{-\frac34}f'(r)\chi.
\end{align}
Substituting this into the Lagrangian neatly absorbs the problematic spin-connection term along the temporal direction, yielding:
\begin{align}
    \mathcal L_f = \chi^\dagger \Bigg( &i\partial_t +i\sigma^x\left[f(r)\left(\partial_r+iA_r\right)+\frac12 f'(r)\right] -m\sqrt{f(r)}\,\sigma^z \Bigg)\chi.
\end{align}
The canonical momentum conjugate to $\chi$ is simply $\Pi = i\chi^\dagger$. Performing the Legendre transformation produces the fermionic Hamiltonian:
\begin{align}
    H_f = \int_{r_h}^{\infty} dr\; \chi^\dagger \Bigg( &-i\sigma^x\left[f(r)\left(\partial_r+iA_r\right)+\frac12 f'(r)\right] +m\sqrt{f(r)}\,\sigma^z  \Bigg)\chi.
\end{align}
This expression can be written more compactly by defining a generalized covariant derivative operator with respect to an arbitrary function $a=a(r)$ (see Appendix~\ref{app:cont-diff-inv} for details):
\begin{align}\label{eq:CovDer}
    i\nabla_r^{(a)} \equiv i\left[ a(r)\left(\partial_r+iA_r\right)+\frac12a'(r) \right].
\end{align}
The total fermionic Hamiltonian is then:
\begin{align}
\label{eq:Accel_Hamiltonian}
    H_f = -\int_{r_h}^{\infty} dr\;\chi^\dagger\sigma^x i\nabla_r^{(f)}\chi  +m\int_{r_h}^{\infty} dr\;\sqrt{f(r)}\,\chi^\dagger\sigma^z\chi
\end{align}
The canonical anti-commutation relations for these continuum fields are
\begin{align}
    \{\chi(r),\chi(s)^\dagger\}=\delta(r-s).
\end{align}
To incorporate the gauge sector, we evaluate the Maxwell action:
\begin{align}
    S_g = -\frac{1}{4e^2}\int d^2x\;F_{\mu\nu}F^{\mu\nu}.
\end{align}
Note that $F^{tr}=-F_{tr}$. In the temporal gauge, the electric field is:
\begin{align}
    E = \frac{\partial\mathcal L}{\partial\partial_tA_r} = \frac{1}{e^2}F_{tr} = -\frac{1}{e^2}F^{tr}.
\end{align}
The gauge Hamiltonian is
\begin{align}
    H_g = \frac{e^2}{2}\int_{r_h}^{\infty} dr\;E^2
\end{align}
Finally, varying the action with respect to $A_t$ yields Gauss's law,
\begin{align}
    \partial_rE = j^t.
\end{align}
Using the rescaled fermions, the charge density simplifies exactly to $j^t = \chi^\dagger\chi$, leaving the constraint in the standard flat-space form:
\begin{align}
    \partial_rE=\chi^\dagger\chi.
\end{align}

\paragraph{The Lattice Hamiltonian.}
For lattice simulations, we discretize the spatial coordinate into a lattice of $N$ sites with spacing $a$. The discretization of the kinetic terms requires care. The elegant repackaging in terms of the covariant-derivative \eqref{eq:CovDer} allows us to covariantly discretize the full kinetic Hamiltonian in one shot. It is easy to see this operator is Hermitian, i.e. $\bra{h}i\nabla_r^{(f)}\ket{h}^*=\bra{h}i\nabla^{(f)}_r\ket{h}$.
In terms of this operator, the kinetic and spin/gauge connection terms in the Hamiltonian become
\begin{align}\label{eq:H_kin_spin}
    H_{\rm kin+spin} = -\int_{r_h}^{\infty} dr\;\chi^\dagger i \sigma^x \nabla_r \chi.
\end{align}
Now we discretize in the canonical fashion treating both $f_{n+\frac12} = f(r_n + \frac a2)$ and $U_n = e^{iaA_n}$ as link variables, which yields the following anti-Hermitian derivative operator -- as required by hermiticity of the Hamiltonian:
\begin{align}\label{eq:cov-der}
    \nabla_{n, m} &= \frac{f_{n+\frac12}U_n\delta_{n+1, m} - f_{n-\frac12} U_{n-1}^\dagger \delta_{n-1, m}}{2a}.
\end{align}
As a sanity check, consider:
\begin{align}\label{eq:cont-lim-cov-der}
    \nabla_{n, m}\chi_m &= \frac1{2a}\Bigg[\left(f(r_n)+\frac{a}{2}f'(r_n)\right)(1+iaA_r(r_n))\left(\chi(r_n)+a\chi'(r_n)\right)\nonumber\\
    &\qquad-\left(f(r_n)-\frac{a}{2}f'(r_n)\right)(1-iaA_r(r_n))\left(\chi(r_n)-a\chi'(r_n)\right)\Bigg]+\dots\nonumber\\
    &=f\left(\partial_r + iA_r\right)\chi + \frac12 f'\chi + O(a^2).
\end{align}
Hence, in the continuum limit, we indeed reproduce the Hamiltonian of equation \eqref{eq:H_kin_spin}. The discretization yields:
\begin{align}
    H_{\rm kin+spin} = -\frac{i}{2a}\sum_{n=0}^{N-2}f_{n+\frac12}\left[\chi_n^\dagger U_n\chi_{n+1} - \chi_{n+1}^\dagger U_n^\dagger \chi_n\right].
\end{align}
where the lattice fermions are related to the continuum ones via $\chi_n = {\sqrt{a}}\chi(r_n)$ and they obey:
\begin{align}
    \{\chi_n, \chi_m^\dagger\} = \delta_{n,m}\,.
\end{align}
The mass term maps to the standard Kogut-Susskind staggered mass, $\chi^\dagger\sigma^z\chi \rightarrow (-1)^n\chi_n^\dagger\chi_n$. The full fermionic Hamiltonian becomes:
\begin{align}
    H_f & = -\frac{i}{2a}\sum_{n=0}^{N-2}f_{n+\frac12}\left[\chi_n^\dagger U_n\chi_{n+1} - \chi_{n+1}^\dagger U_n^\dagger \chi_n\right]+ m\sum_{n=0}^{N-1}(-1)^n\sqrt{f_n}\chi_n^\dagger \chi_n  .
\end{align}
We impose open boundary conditions and gauge fix the spatial links to unity, i.e. $U_n=1$. The discretized Gauss law reads: 
\begin{align}
E_{n+1}-E_n&=Q_n, & Q_n & = \chi_n^\dagger \chi_n - \frac{1-(-1)^n}{2},
\end{align} 
and may be solved recursively. Setting the boundary electric field $E_0=0$, the local electric field becomes $E_n = \sum_{j=0}^n Q_j$. The gauge Hamiltonian $H_E$ then explicitly couples all charges via a long-range Coulomb interaction:
\begin{align}
    H_E =\frac{e^2a}{2}\sum_{n=0}^{N-1}E_n^2 = \frac{e^2a}{2} \sum_{n=0}^{N-1} \left( \sum_{j=0}^nQ_j \right)^2.
\end{align}
The completely gauge-fixed fermionic Hamiltonian is:
\begin{align}
    H_f = &-\frac{i}{2a}\sum_{n=0}^{N-2} f_{n+\frac12} \left[ \chi_n^\dagger\chi_{n+1} - \chi_{n+1}^\dagger\chi_n \right] +m\sum_{n=0}^{N-1}(-1)^n\sqrt{f_n}\,\chi_n^\dagger\chi_n .
\end{align}

\paragraph{The Qubit Hamiltonian.}
To map the fermions to qubits for our simulations, we apply the standard Jordan-Wigner transformation. The fermionic creation and annihilation operators are mapped to Pauli matrices $(X, Y, Z)$ via:
\begin{align}\label{eq:JW-convention}
    \chi_n &= \frac{X_n-iY_n}{2} \prod_{m=0}^{n-1} Z_m, & \chi_n^\dagger &= \frac{X_n+iY_n}{2} \prod_{m=0}^{n-1} Z_m.
\end{align}
The staggered charge is:
\begin{align}
    \chi_n^\dagger\chi_n &= \frac{1+Z_n}{2}, & Q_n &= \frac{(-1)^n+Z_n}{2}.
\end{align}
The nearest-neighbor hopping reduces to local two-body interactions:
\begin{align}
    -i\left[ \chi_n^\dagger\chi_{n+1} - \chi_{n+1}^\dagger\chi_n \right] &= \frac12 \left( X_nY_{n+1}-Y_nX_{n+1} \right).
\end{align}
Summing the fermionic and electric components, the final spin Hamiltonian for the Schwarzschild frame is:
\begin{align}
    H = &\frac{1}{4a}\sum_{n=0}^{N-2} f_{n+\frac12} \left( X_nY_{n+1}-Y_nX_{n+1} \right) \nonumber
 +\frac{m}{2}\sum_{n=0}^{N-1} (-1)^n\sqrt{f_n}\left(1+Z_n\right) \\&\hspace{15em} +\frac{e^2a}{2} \sum_{n=0}^{N-1} \left( \sum_{j=0}^n \frac{(-1)^j+Z_j}{2} \right)^2.
\end{align}

A few remarks are in order. Setting $e^2=0$ freezes the gauge field and the electric term drops out, leaving a free staggered-fermion Hamiltonian on AdS$_2$. This is the setting considered by \cite{ikeda2025quantumsimulationfermionsads2, ikeda2026geometryinducedchiraltransport}, but our Hamiltonian does not reduce to theirs: in a curved background the kinetic term and the spin-connection term are each separately non-Hermitian and become Hermitian only in combination, and discretizing the two independently, as done there, is not covariant. By packaging both terms into the single generalized covariant derivative \eqref{eq:CovDer} before discretizing, we instead obtain the manifestly anti-Hermitian difference operator \eqref{eq:cov-der}, so that the spin connection is incorporated covariantly and the Hamiltonian is Hermitian at every finite $a$ while reproducing the correct continuum limit \eqref{eq:cont-lim-cov-der}. For $e^2\neq0$ we further generalize this program to include a dynamical $\mathrm{U}(1)$ gauge field, introduced through the link variables $U_n=e^{iaA_n}$ and the electric energy $\tfrac{e^2a}{2}\sum_n E_n^2$, with Gauss's law solved to give the long-range Coulomb interaction above.

\subsection{Global AdS$_2$ Hamiltonian: continuum and lattice}
\label{sec:GlobalAdS}
We now describe QED$_2$ in the global AdS$_2$ frame \eqref{eq:globalAdS}, which corresponds to a constant dilaton in JT gravity. Because no Killing direction is preferred, the $\mathrm{SL}(2,\mathbb{R})$ symmetry of the background is preserved and the natural vacuum is the $\mathrm{SL}(2,\mathbb{R})$-invariant state. We construct the continuum Hamiltonian via a Legendre transform on a constant $\tau$-slice, discretize using staggered fermions, and apply a Jordan-Wigner transformation to obtain the qubit Hamiltonian.

\paragraph{The Continuum Hamiltonian.}

The non-vanishing zweibein components are
\begin{align}
    e^0 &= L\cosh\rho \, d\tau & e^1 &= L \, d\rho
\end{align}
which yield a metric determinant $\sqrt{-g} = L^2\cosh\rho$. Solving the torsion-free Maurer-Cartan equation $de^0+\omega^{01}\wedge e_1=0$ yields the single non-zero spin connection component:
\begin{align}
    \omega_\tau^{01} = \sinh\rho.
\end{align}
The fermionic Lagrangian in the temporal gauge $A_\tau = 0$ is constructed using the curved-space Dirac matrices $\Gamma^\tau = \frac{1}{L\cosh\rho}\gamma^0$ and $\Gamma^\rho = \frac{1}{L}\gamma^1$:
\begin{align}
    \mathcal{L}_f = \sqrt{-g} \bar\psi \left( i\Gamma^\tau D_\tau + i\Gamma^\rho D_\rho - m \right) \psi \,,
\end{align}
where the covariant derivatives are $D_\tau = \partial_\tau + \frac{1}{2}\sinh\rho \, \sigma^x$ and $D_\rho = \partial_\rho + iA_\rho$. We redefine the fermion field as $\chi = \sqrt{L} \psi$.  Then the kinetic and spin-connection terms elegantly combine with the spatial derivative, yielding the fermionic Hamiltonian:
\begin{align}
    H_f = \int d\rho \; \chi^\dagger \left[ -i\sigma^x \left( \cosh\rho (\partial_\rho + iA_\rho) + \frac{1}{2}\sinh\rho \right) + m L\cosh\rho \sigma^z \right] \chi\,.
\end{align}
As before, we use the covariant derivative $i\nabla_\rho^{(\cosh)} \equiv i\left[ \cosh\rho (\partial_\rho + iA_\rho) + \frac{1}{2}\sinh\rho \right]$, we can write this in the compact form:
\begin{align}
\label{eq:inertial_Hamiltonian}
    H_f = -\int d\rho \; \chi^\dagger \sigma^x i\nabla_\rho^{(\cosh)} \chi + \int d\rho \; L\cosh\rho  m \chi^\dagger \sigma^z \chi \,.
\end{align}
In the gauge sector, the Maxwell action $S_g = -\frac{1}{4e^2} \int d\tau d\rho \sqrt{-g} F_{\mu\nu} F^{\mu\nu}$ produces the conjugate momentum $E = \frac{1}{e^2 L^2 \cosh\rho} \partial_\tau A_\rho$. This yields the Hamiltonian:
\begin{align}
    H_g = \frac{e^2 L^2}{2} \int d\rho \cosh\rho E^2\,.
\end{align}
Gauss's law reduces exactly to the standard flat-space form, 
\begin{align}
    \partial_\rho E = \chi^\dagger \chi.
\end{align}

\paragraph{The Lattice Hamiltonian.}

The spatial coordinate $\rho$ is mapped onto a lattice of $N$ sites with spacing $a$,  staggered fermions $\chi_n = \sqrt{a} \chi(\rho_n)$. Using the link variables $U_n = e^{iaA_n}$, the covariant derivative maps to the finite-difference operator:
\begin{align}
    -\int d\rho \; \chi^\dagger \sigma^x i\nabla_\rho^{(\cosh)} \chi \longrightarrow -\frac{i}{2a} \sum_{n=0}^{N-2} \cosh\rho_{n+\frac12} \left[ \chi_n^\dagger U_n \chi_{n+1} - \chi_{n+1}^\dagger U_n^\dagger \chi_n \right]\,.
\end{align}
Enforcing open boundary conditions and gauge-fixing $U_n = 1$, we define the staggered charge $Q_n = \chi_n^\dagger \chi_n - \frac{1-(-1)^n}{2}$. Solving Gauss's law with $E_0 = 0$ yields $E_n = \sum_{j=0}^n Q_j$. Summing all sectors gives the gauge-fixed lattice Hamiltonian:
\begin{align}
    H_f = &-\frac{i}{2a} \sum_{n=0}^{N-2} \cosh\rho_{n+\frac12} \left[ \chi_n^\dagger \chi_{n+1} - \chi_{n+1}^\dagger \chi_n \right] + m L \sum_{n=0}^{N-1} (-1)^n \cosh\rho_n \chi_n^\dagger \chi_n \nonumber\\
    &\hspace{15em}+ \frac{e^2 L^2 a}{2} \sum_{n=0}^{N-1} \cosh\rho_n \left( \sum_{j=0}^n Q_j \right)^2\,.
\end{align}

\paragraph{The Qubit Hamiltonian.}

Finally, we apply the Jordan-Wigner transformation, introduced in \eqref{eq:JW-convention}, $\chi_n^\dagger \chi_n = \frac{1+Z_n}{2}$ and map the staggered nearest-neighbor hopping to local Pauli interactions. The resulting qubit Hamiltonian for the global $\text{AdS}_2$ inertial frame is:
\begin{align}
    H = &\frac{1}{4a} \sum_{n=0}^{N-2} \cosh\rho_{n+\frac12} \left( X_n Y_{n+1} - Y_n X_{n+1} \right) + \frac{m L}{2} \sum_{n=0}^{N-1} (-1)^n \cosh\rho_n (1+Z_n) \nonumber\\
    &\hspace{15em} + \frac{e^2 L^2 a}{2} \sum_{n=0}^{N-1} \cosh\rho_n \left( \sum_{j=0}^n \frac{(-1)^j + Z_j}{2} \right)^2\,.
\end{align}

\section{Screening in massless QED$_2$ on AdS$_2$: zero and finite temperature}\label{sec:Screening}
In this section, we determine whether massless QED$_2$ on AdS$_2$ screens or confines at both zero and finite temperature. Our diagnostic is the static potential, which we compute in the continuum by
bosonization. At zero temperature we work on pure AdS$_2$ (no black hole) and distinguish two natural vacua: the Boulware vacuum (i.e. the ground state of the Schwarzschild-frame Hamiltonian \eqref{eq:Accel_Hamiltonian}) and the $\mathrm{SL}(2,\mathbb{R})$-invariant vacuum (i.e. the ground state of the global AdS$_2$ Hamiltonian \eqref{eq:inertial_Hamiltonian}). 

The static potential in this setting has a confusing history. Refs.~\cite{Alimohammadi:2000fg, MohseniSadjadi:2000mg, MohseniSadjadi:2000nt} identify the ground-state energy in the presence of the probe charges directly with the static potential, and on this basis reach conclusions that depend on how the charges are taken to infinity. For instance, \cite{Alimohammadi:2000fg}
finds a potential growing with the geodesic separation in dS$_2$ and reads off
confinement, whereas \cite{MohseniSadjadi:2000nt} reports confinement when
$r_2\to\infty$ at fixed $r_1$ but screening when $r_1\to0$ at fixed $r_2$, even
though both limits send the geodesic separation $d=L\log(r_2/r_1)$ to infinity;
\cite{MohseniSadjadi:2000mg} similarly finds region-dependent behavior. The claims of confinement in curved space \cite{Alimohammadi:2000fg, MohseniSadjadi:2000mg, MohseniSadjadi:2000nt} are especially striking in the massless case, given that the massive model is expected to become confining as $m\rightarrow\infty$. 

We argue that the ground-state energy should not be identified with the binding potential: in curved space the probe self-energies are position dependent (unlike the position-independent constants of flat space) and must be subtracted
before the potential can be read off. Once this is done, we find a single, screened phase, independent of how $d\to\infty$ is taken, in agreement with the one-form symmetry expectation described in Sec.~\ref{sec:one-form-sym}. Since the massless model is exactly solvable, we  restrict throughout this section to the massless Schwinger model. The massive model, unlike the massless model, is not exactly solvable, and is addressed in Section~\ref{sec:Simulation} via Hamiltonian simulation, where the screened phase is validated in a real-time simulation of the string-breaking process.

\subsection{Constraints from the explicitly broken one-form symmetry of QED$_2$}\label{sec:one-form-sym}

An equivalent characterization of confinement can be given as the presence of an unbroken electric one-form symmetry.
The phase of the IR theory can be constrained using the $\mathrm{U}(1)$ electric one-form symmetry of two-dimensional pure Maxwell theory, which is explicitly broken upon coupling to dynamical charged fermions~\cite{Gaiotto:2014kfa,Tong:GaugeTheory}. It is instructive to begin with the (classically) confining pure Maxwell theory, where this symmetry is present. Switching off the dynamical fermions, the equation of motion reads
\begin{align}
    \nabla_\mu F^{\mu\nu}=0.
\end{align}
This equation expresses the conservation of the electric one-form symmetry current $F^{\mu\nu}$, which is a two-form. Since no charged matter is available on which a Wilson line can end, electric flux cannot be screened and the symmetry remains unbroken. In the generalized-symmetry language, this is reflected in an area law for Wilson loops and the presence of stable electric flux tubes~\cite{Gaiotto:2014kfa,Tong:GaugeTheory}.

Gauss's law fixes a constant electric field between the charges, $E(r)=q$ for $r_1<r<r_2$ and $E=0$ otherwise, exactly as in flat space, and the entire energy resides in the connecting flux tube,
\begin{align}
    V_{\rm Maxwell}(r_1,r_2)=\frac{e^2}{2}\int_0^\infty dr\,E^2=\frac{e^2q^2}{2}\,(r_2-r_1).
\end{align}
This grows without bound as the charges are separated. The potential implies a constant string tension $\sigma=e^2q^2/2$. This is analogous to the unbounded growth of the confining potential in flat space. There is no screening cloud: an isolated probe sources a flux tube reaching the boundary and costs infinite energy, the hallmark of confinement.

Dynamical fermions alter this picture. They source the gauge field,
\begin{align}
    \nabla_\mu F^{\mu\nu}=j^\nu\neq0,
\end{align}
and thereby explicitly break the electric one-form symmetry~\cite{Gaiotto:2014kfa}. Equivalently, Wilson lines carrying the charge of the dynamical fermions can terminate on local charged operators, so the corresponding electric flux tubes are no longer protected against breaking. The absence of an exact one-form symmetry does not by itself prove screening, but it removes the symmetry-based constraint and permits integer-charge probes to be screened by the dynamical matter~\cite{Gaiotto:2014kfa,Tong:GaugeTheory}.

In the massless Schwinger model, one finds that vacuum polarization generates a mass for the gauge field and screens external charges~\cite{Schwinger:1962tp,Coleman:1975pw,Coleman:1976uz}. Moreover, we show that this phenomenon persists in curved space. As the explicit calculation to follow confirms, the fermions endow each probe with a finite, position-dependent self-energy. Once these individual self-energies are subtracted, the binding potential saturates rather than rising without bound, consistently with the explicit breaking of the electric one-form symmetry.

\subsection{Schwarzschild AdS$_2$ at zero temperature}

In this subsection, we evaluate the static potential of a pair of equal and opposite probe charges in the pure AdS$_2$ background in Schwarzschild coordinates at zero temperature. We proceed by bosonizing the massless theory with boundary conditions \eqref{eq:BC} that enforce the vanishing of the electric field at the boundaries, which allows us to get a continuum energy functional in the presence of probe charges. Extremizing this functional, we obtain a semi-classical solution, whose energy we evaluate on-shell. Upon carefully subtracting the position-dependent self-energy contributions, we obtain the static potential \eqref{eq:StatPot}. Following this, we use the semi-classical solution to calculate the electric flux profile explicitly in \eqref{eq:ElecProfileSchwarzschild}. Taken together, this demonstrates that the theory is screened rather than confined, in agreement with the one-form symmetry considerations of Section~\ref{sec:one-form-sym}. 

\paragraph{Derivation of the bosonized Hamiltonian.}Consider again the fermionic Lagrangian:\footnote{In tortoise coordinates, $dx = dr f(r)^{-1}$ and $\partial_x = f(r)\partial_r$. This implies
\begin{align}
    S_f = \int dxdt\sqrt{f(x)}\;\psi^\dagger\Bigg(i\left(\partial_t+iA_t\right)  &+i\sigma^x\left(\partial_x+iA_x\right) + i\frac{f'(x)}{4}\sigma^x - m\sqrt{f(x)}\sigma^z\Bigg) \psi\nonumber
\end{align}
Introduce a rescaled fermionic variable $\zeta(x)  = f(x)^{\frac14}\psi(x)$. A careful computation with the product rule on $\partial_x(f(x)^{-\frac14}\zeta(x))$ yields a cancellation between the spin connection and gauge-covariant derivative, which yields:
\begin{align}
    S_f &= \int dxdt\;\bar\zeta\left[i\gamma^0D^{\rm flat}_t + i\gamma^0D_x^{\rm flat} - m\sqrt{f(x)}\sigma^z \right]\zeta & D_\mu^{\rm flat} &=\partial_{\mu} + iA_{\mu}
\end{align}
For our current purposes, we focus on the massless case $m=0$. In tortoise coordinates, the theory looks like the flat space Schwinger model. This is why \cite{MohseniSadjadi:2000mg} does not explicitly write a spin connection term.}
\begin{align}
    \mathcal{L}_f = \psi^\dagger\frac{i}{\sqrt{f(r)}}\left(\partial_t +i\sqrt{f(r)}\sigma^x(\partial_r+iA_r) + i\frac{f'(r)}{4\sqrt{f(r)}}\sigma^x  - m\sigma^z\right) \psi\,.
\end{align}
We will bosonize this Lagrangian using the standard dictionary:
\begin{align}
    j^{\mu} = \bar\psi\Gamma^{\mu}\psi\xleftrightarrow{}j^{\mu} = \frac{1}{\sqrt{\pi}}\varepsilon^{\mu\nu}\partial_\nu\phi\,,
\end{align}
where $\varepsilon^{\mu\nu} = \epsilon^{\mu\nu}/\sqrt{-g}$ with $\epsilon^{tr}=+1$. We focus on the massless model, i.e. $m = 0$. Since we established that the above Lagrangian is covariant, the fully gauge-fixed continuum Hamiltonian is simply the covariant kinetic term where we focus on the static case with $\partial_\tau \phi=0$:
\begin{align}
    S_f =\frac12\int d^2x\sqrt{-g}\;g^{rr}(\partial_r\phi)^2.
\end{align}
The Gauss law is modified via an external charge:
\begin{align}
    \partial_r E &= j^t + \rho_0(r) = \frac{1}{\sqrt{\pi}}\partial_r\phi +\partial_r\theta &\rho_0(r) = \partial_r\theta(r)& = q_1\delta(r-r_1 )+q_2\delta(r-r_2)\,.
\end{align}
Integrating this constraint, we obtain $E = \frac{1}{\sqrt{\pi}}\phi + \theta$. The full Hamiltonian is 
\begin{align}
    H & = \int dr\left[\frac{e^2}{2}\left(\frac{\phi}{\sqrt{\pi}} + \theta\right)^2 + \frac{f(r)}{2}(\partial_r\phi)^2\right]\,.
\end{align}
Taking the variation $\phi\rightarrow\phi+\delta\phi$ followed by integration by parts yields
\begin{align}
    \delta H&= \int dr\left[\frac{e^2}{\sqrt{\pi}}\left(\frac{\phi}{\sqrt{\pi}} + \theta\right) - \partial_r\left(f(r)\partial_r\phi(r)\right)\right]\delta\phi\,.
\end{align}
Hence, the equation of motion is
\begin{align}\label{eq:EOM-Sch}
    -\partial_r\left(f(r)\partial_r\phi(r)\right)+\frac{e^2}{\pi}\phi = -\frac{e^2}{\sqrt{\pi}}\theta\,.
\end{align}
\paragraph{Solution for two external charges.}
In the vacuum with no static charges $\theta=0$ with $f(r) = r^2/L^2$, this yields the differential equation
\begin{align}
    r^2\frac{d^2\phi}{dr^2}+2r\frac{d\phi}{dr} - \frac{e^2L^2}{\pi}\phi = 0\,.
\end{align}
The Green's function for this equation is
\begin{align}
    G(r, r') &= \frac{1}{2\nu+1}r_<^\nu r_>^{-\nu-1} &\nu&= -\frac{1}{2}+\sqrt{\frac
    14 + \frac{e^2L^2}{\pi}}\,,
\end{align}
where $r_< = \min\{r, r'\}$ and $r_> = \max\{r, r'\}$. Explicitly, we have
\begin{align}
    \left[-r^2\partial_r^2-2r\partial_r+\nu(\nu+1)\right]G(r,r') = \delta(r-r')\,.
\end{align}
In the presence of sources, following \cite{Gross:1995bp}, we impose boundary conditions that enforce vanishing of the electric field at both ends:
\begin{align}\label{eq:BC}
    E(0) &= E(\infty) = 0&\iff& &\phi(0) &= 0, &\phi(\infty) &= -\sqrt{\pi} (q_1+q_2),
\end{align}
where we have chosen the potential for the probe density as:
\begin{align}
    \theta(r) =q_1\Theta(r-r_1) +q_2\Theta(r-r_2) = \begin{cases}
        0, &r<r_1\\
        q_1, &r_1<r<r_2\\
        q_1+q_2, &r_2<r<\infty
    \end{cases}.
\end{align}
The solution to the inhomogeneous differential equation is gotten simply by integrating the charge against the Green's function:
\begin{align}\label{eq:sol}
    \phi_0(r) &= - \frac{e^2L^2}{\sqrt{\pi}}\int_{0}^\infty dr'\; G(r, r')\theta(r')\nonumber\\&
=
-\sqrt{\pi}
\begin{cases}
\displaystyle
\frac{\nu+1}{2\nu+1}
\left[
q_1\left(\frac{r}{r_1}\right)^\nu
+
q_2\left(\frac{r}{r_2}\right)^\nu
\right],
& 0<r<r_1,
\\[1.3em]
\displaystyle
q_1
\left[
1-\frac{\nu}{2\nu+1}
\left(\frac{r_1}{r}\right)^{\nu+1}
\right]
+
q_2
\frac{\nu+1}{2\nu+1}
\left(\frac{r}{r_2}\right)^\nu,
& r_1<r<r_2,
\\[1.3em]
\displaystyle
q_1
\left[
1-\frac{\nu}{2\nu+1}
\left(\frac{r_1}{r}\right)^{\nu+1}
\right]
+
q_2
\left[
1-\frac{\nu}{2\nu+1}
\left(\frac{r_2}{r}\right)^{\nu+1}
\right],
& r>r_2.
\end{cases}
\end{align}
\paragraph{The on-shell energy.} To compute the vacuum energy, we simply plug this solution into the on-shell energy functional:
\begin{align}
    H[\phi_0; q_1, q_2] &= \int dr\left[\frac{e^2}{2}\left(\frac{\phi_0}{\sqrt{\pi}} + \theta\right)^2 + \frac{f(r)}{2}(\partial_r\phi_0)^2\right]\nonumber\\
    &=\int dr\;\frac{e^2}{2}\theta^2 -\frac{e^4L^2}{2\pi}\int dr\,dr'\;\theta(r)G(r, r')\theta(r').
\end{align}
It is straightforward to evaluate the integral, assuming $r_1<r_2$:
\begin{equation}
\begin{aligned}
H[\phi_0;q_1, q_2]
= \frac{e^2}{2(2\nu+1)}
\Biggl[&
q_1^2r_1
+ 2q_1 q_2 r_1
\left(\frac{r_1}{r_2}\right)^\nu + q_2^2 r_2
\Biggr].
\end{aligned}
\end{equation}
There are two important limits of this equation that we now highlight. 
\begin{enumerate}
    \item {$(q_1,q_2)=(q,-q)$: the neutral pair.} 
    \begin{align}
    \label{eq:TotalEnergy}
        H[\phi_0; q, -q] = \frac{e^2q^2}{2(2\nu+1)}\left[r_2+r_1\left\{1-2\left(\frac{r_1}{r_2}\right)^\nu\right\}\right]\,.
    \end{align}
    \item {$(q_1,q_2)=(q,0)$: the single-charge self-energy located at $r = r_0$.} 
    \begin{align}\label{eq:SelfEnergy}
         H[\phi_0;q,0] = H[\phi_0;0,-q] = \frac{e^2q^2}{2(2\nu+1)}r_0.
    \end{align}
\end{enumerate}

\paragraph{The $q\bar q$-potential.} We are now ready to write down the $q\bar q$ potential of a pair of equal and opposite charges, which is obtained from the total vacuum energy by subtracting the self-energy contributions due to the two probe charges. 
\begin{align}\label{eq:StatPot}
    V_{q\bar q}(r_1, r_2) & = H[\phi_0; q, -q] - H[\phi_0; q,0]- H[\phi_0; 0,q] = -\frac{e^2q^2}{2\nu+1}\left(\frac{r_1}{r_2}\right)^\nu r_1.
\end{align}
This expression is a hallmark of a screened phase, and warrants several comments. 
\begin{enumerate}
    \item The geodesic separation between the two probe charges is given by
    \begin{align}
        d = L\log\frac{r_2}{r_1}.
    \end{align}
    When $r_1\rightarrow0$ with $r_2$ held fixed, $d\rightarrow\infty$: in this limit the potential vanishes polynomially in $r_1$ and exponentially in $d$. On the other hand, when $r_2\rightarrow\infty$ with $r_1$ fixed, we still have $d\rightarrow\infty$, leading to a polynomial vanishing in $r_2$ and an exponential one in $d$. Explicitly:
    \begin{align}
        V_{q\bar q}(r_1, r_2) &= \begin{cases}
            -\frac{e^2q^2}{2\nu+1}r_1 e^{-\nu \frac d L},&r_1\text{ fixed}\\
             -\frac{e^2q^2}{2\nu+1}r_2 e^{-(\nu+1) \frac d L},&r_2\text{ fixed}
        \end{cases}
    \end{align}
    In either regime, the massless theory is \textit{screened}.
    \item In Refs.~\cite{Alimohammadi:2000fg, MohseniSadjadi:2000mg, MohseniSadjadi:2000nt}, the massless Schwinger model in curved space is said to be confining based on the fact that the contributions to the vacuum energy coming from \eqref{eq:SelfEnergy} diverge linearly as $r\rightarrow\infty$. However, this linear growth should not be interpreted as a signature of confinement since it occurs even for a single probe charge and comes from the electrostatic self-energy of each of the probe charges. This dependence of the self-energy of a single probe charge on the position reflects the absence of translational symmetry under shifts of the radial coordinate, and is not reflective of confinement.     The statement of confinement concerns the binding potential of a pair of equal and opposite probe charges, which, in this case, is decaying rather than growing. This is precisely why we subtract the self-energy contributions $H[\phi_0;q,0]+H[\phi_0;0,-q]$ from the vacuum energy to obtain the true static $q\bar q$-potential. As we explain shortly, this is also reflected in the electric field profile.
    
    \item In contrast, the flat space single-probe self-energies are translationally invariant, and hence constants. This is precisely why they are not typically dropped from the vacuum energy, which is directly identified in flat space as the $q\bar q$ potential. We analyze this in depth in Section~\ref{sec:flatspace}.
\end{enumerate}
\paragraph{The electric field profile.}
We now specialize to a equal and opposite pair of external charges,
\begin{align}
    q_1=-q_2=q,
\end{align}
For this configuration, using \eqref{eq:sol} and
\begin{align}
    E(r)=\frac{\phi_0(r)}{\sqrt{\pi}}+\theta(r),
\end{align}
the electric field profile is
\begin{align}
E(r)=
\begin{cases}
\displaystyle
-\frac{e^2L^2}{\pi(2\nu+1)}\,q\,
\frac{r^\nu}{\nu}
\left(r_1^{-\nu}-r_2^{-\nu}\right),
& 0<r<r_1,
\\[1.4em]
\displaystyle
q-\frac{e^2L^2}{\pi(2\nu+1)}\,q
\left[
\frac{1}{\nu+1}
\left(1-\frac{r_1^{\nu+1}}{r^{\nu+1}}\right)
+
\frac{1}{\nu}
\left(1-\frac{r^\nu}{r_2^\nu}\right)
\right],
& r_1<r<r_2,
\\[1.4em]
\displaystyle
-\frac{e^2L^2}{\pi(2\nu+1)}\,q\,
\frac{r^{-\nu-1}}{\nu+1}
\left(r_2^{\nu+1}-r_1^{\nu+1}\right),
& r>r_2.
\end{cases}
\end{align}
\begin{figure}[t]
    \centering
    \includegraphics[width=0.5\linewidth]{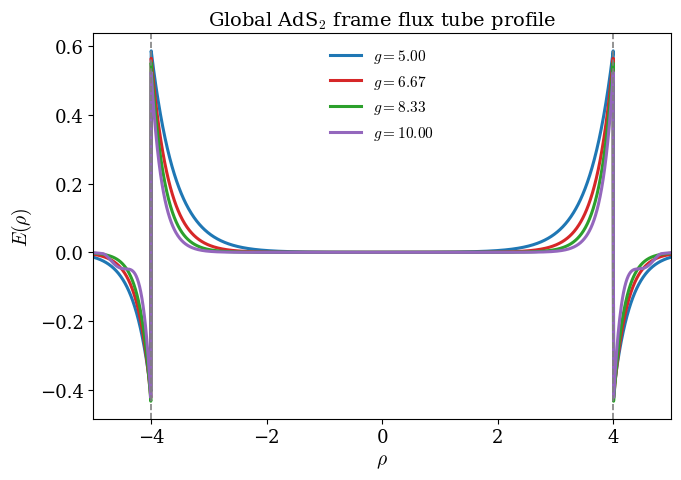}\includegraphics[width=0.5\linewidth]{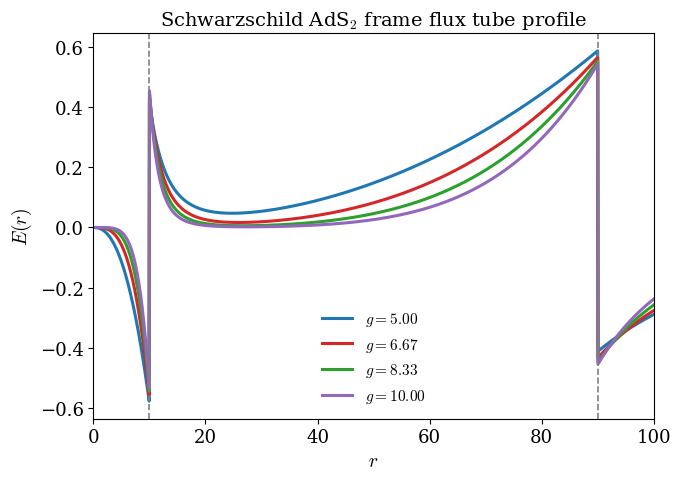}
    \caption{Electric field profile for a neutral pair of external charges \(q_1=-q_2=q\), plotted for several values of the gauge coupling \(g\): global AdS\(_2\), \(E(\rho)\) (left), and the Schwarzschild frame, \(E(r)\) (right). The dashed vertical lines mark the charge insertions at \(\rho=\rho_{1,2}\) (resp. \(r=r_{1,2}\)). Dynamical massless fermions screen the flux tube, producing tails outside the interval between the charges; in the weak-coupling limit the profile approaches the sharp constant background field between the charges, while increasing \(g\) smooths it, and when \(L/\nu\ll d\) the field for \(r\in[r_1,r_2]\) is exponentially small in \(g\). Unlike the Schwarzschild-frame profile, the global profile is not a function only of the geodesic separation between the probes, because the global Hamiltonian contains the redshift factor \(\cosh\rho\).}
    \label{fig:flux-tube-profile}
\end{figure}
This form can be written more geometrically in terms of the radial geodesic
separation
\begin{align}
    D(a,b)=L\log\frac{b}{a},
    \qquad
    d=D(r_1,r_2).
\end{align}
Equivalently, the profile is
\begin{align}\label{eq:ElecProfileSchwarzschild}
E(r)
=
\begin{cases}
\displaystyle
-\frac{\nu+1}{2\nu+1}
\left[
q e^{-\nu D(r,r_1)/L}
-
q e^{-\nu D(r,r_2)/L}
\right],
& 0<r<r_1,
\\[1.4em]
\displaystyle
\frac{\nu}{2\nu+1}
q e^{-(\nu+1)D(r_1,r)/L}
+
\frac{\nu+1}{2\nu+1}
q e^{-\nu D(r,r_2)/L},
& r_1<r<r_2,
\\[1.4em]
\displaystyle
\frac{\nu}{2\nu+1}
\left[
q e^{-(\nu+1)D(r_1,r)/L}
-
q e^{-(\nu+1)D(r_2,r)/L}
\right],
& r>r_2.
\end{cases}
\end{align}
This makes explicit that the flux tube profile is controlled by proper radial distances rather than coordinate separations. Outside the interval between the charges, the electric field decays exponentially with the geodesic distance from the nearest insertion, while the overall amplitude is set by the proper separation $d=D(r_1,r_2)$ of the neutral pair. Inside the interval \(r_1<r<r_2\), the field contains the direct contribution \(q\) from the background charge together with the induced response of $\phi_0$, which decreases exponentially as $g$ is raised since $\nu\sim gL/\pi+\dots$ at strong coupling, unlike what one expects in a confining phase.

\subsection{Global AdS$_2$ at zero temperature}
Having analyzed the problem of confinement in the Schwarzschild frame, we now turn to the analogous problem in the global AdS$_2$ frame. The strategy is the same as the preceding section: we derive the bosonized Hamiltonian for massless QED$_2$ on global AdS$_2$, obtain the semi-classical vacuum solution, and use it to explicitly evaluate the static potential of a pair of equal and opposite probe charges \eqref{eq:global_qbarq_potential} and the electric field profile \eqref{eq:global_flux_profile_exact}. We close the subsection with a comparison between the global and Schwarzschild frames.

\paragraph{Derivation of the bosonized Hamiltonian.} We again focus on the massless Schwinger model at zero density, $m=0$. Bosonizing with the same dictionary $j^\mu=\frac{1}{\sqrt\pi}\varepsilon^{\mu\nu}\partial_\nu\phi$ as before and solving Gauss's law in temporal gauge, the reduced Hamiltonian is
\begin{align}
    H
    =
    \frac12
    \int d\rho\,a(\rho)
    \left[
        \Pi_\phi^2
        +
        (\partial_\rho\phi)^2
        +
        \mu^2
        \left(\phi+\sqrt{\pi}\,\theta(\rho)\right)^2
    \right],
    \qquad
    \mu^2\equiv \frac{e^2L^2}{\pi}\,,
\end{align}
where we abbreviate the redshift factor as $a(\rho)\equiv \cosh\rho$. Here $\theta(\rho)$ is the integrated external charge density. For two external charges,
\begin{align}
    \rho_0(\rho)
    =
    q_1\delta(\rho-\rho_1)
    +
    q_2\delta(\rho-\rho_2),
    \qquad
    \rho_1<\rho_2,
\end{align}
we integrate $\partial_\rho\theta=\rho_0$ to obtain
\begin{align}
    \theta(\rho)
    =
    q_1\Theta(\rho-\rho_1)
    +
    q_2\Theta(\rho-\rho_2).
\end{align}
This convention fixes the integration constant so that $\theta(\rho)\to0$ at
the left boundary. The electric field is
\begin{align}
    E(\rho)=\frac{\phi(\rho)}{\sqrt{\pi}}+\theta(\rho).
\end{align}
We impose that the electric field vanishes at both global boundaries, which is equivalent to
\begin{align}
    E(-\infty)=E(+\infty)=0
    \qquad\Longleftrightarrow\qquad
    \phi(-\infty)=0,
    \qquad
    \phi(+\infty)=-\sqrt{\pi}(q_1+q_2).
    \label{eq:global_E_vanishing_BC}
\end{align}
For the static potential, we minimize the Hamiltonian in a time-independent external charge sector. Since the kinetic term is positive and contains no source term, the minimizing configuration has $\Pi_\phi=0$. The relevant static energy functional is therefore
\begin{align}
    H[\phi]=\frac12
    \int d\rho\,a(\rho)
    \left[
        (\partial_\rho\phi)^2
        +
        \mu^2
        \left(\phi+\sqrt{\pi}\,\theta(\rho)\right)^2
    \right].
\end{align}
Taking the variation $\phi\to\phi+\delta\phi$ and integrating by parts gives
\begin{align}
    \delta H
    =
    \int d\rho
    \left[
        -\partial_\rho\left(a(\rho)\partial_\rho\phi\right)
        +
        \mu^2a(\rho)\phi
        +
        \sqrt{\pi}\mu^2a(\rho)\theta(\rho)
    \right]\delta\phi .
\end{align}
Thus the saddle obeys
\begin{align}
    \left[
        -\partial_\rho\left(a(\rho)\partial_\rho\right)
        +
        \mu^2a(\rho)
    \right]\phi_0(\rho)
    =
    -\sqrt{\pi}\mu^2a(\rho)\theta(\rho).
    \label{eq:global_phi0_eom}
\end{align}
After dividing by $a(\rho)=\cosh\rho$, homogeneous version of \eqref{eq:global_phi0_eom} is
\begin{align}
    u''+\tanh\rho\,u'-\mu^2u=0 ,
\end{align}
with
\begin{align}
    \nu
    =
    -\frac12+\sqrt{\mu^2+\frac14}
    =
    -\frac12+\sqrt{\frac14+\frac{e^2L^2}{\pi}}.
\end{align}
A convenient pair of homogeneous solutions is
\begin{align}
    u_R(\rho)
    &=
    (\sech\rho)^{1/2}
    P_{-1/2}^{-(\nu+1/2)}(\tanh\rho),
    &
    u_L(\rho)
    &=
    (\sech\rho)^{1/2}
    P_{-1/2}^{-(\nu+1/2)}(-\tanh\rho),
\end{align}
Here, $P_\alpha^{\beta}(x)$ are the associated Legendre functions. Their asymptotic behavior is
\begin{align}
    u_R(\rho) &\underset{\rho\to+\infty}{\sim} e^{-(\nu+1)\rho},
    &
    u_L(\rho) &\underset{\rho\to-\infty}{\sim} e^{+(\nu+1)\rho}.
\end{align}
Thus $u_R$ decays at the right boundary and $u_L$ decays at the left boundary.
The weighted Wronskian is independent of $\rho$:
\begin{align}
    \mathcal W
    =
    a(\rho)
    \left[
        u_L(\rho)u_R'(\rho)
        -
        u_L'(\rho)u_R(\rho)
    \right]\,.
\end{align}

\paragraph{Solution for two external charges.}
We now solve \eqref{eq:global_phi0_eom} directly for the two-source profile.
Since $\theta$ is constant in each region, the particular solution is constant
in each region. The boundary conditions \eqref{eq:global_E_vanishing_BC} and
continuity of $\phi_0$ and $\partial_\rho\phi_0$ at $\rho_1$ and $\rho_2$ give
\begin{align}
\phi_0(\rho)
=
-\sqrt{\pi}
\begin{cases}
\displaystyle
\frac{u_L(\rho)}{\mathcal W}
\left[
q_1a(\rho_1)u_R'(\rho_1)
+
q_2a(\rho_2)u_R'(\rho_2)
\right],
& \rho<\rho_1,
\\[1.4em]
\displaystyle
q_1
\left[
1+
\frac{a(\rho_1)u_L'(\rho_1)}{\mathcal W}
u_R(\rho)
\right]
+
q_2
\frac{a(\rho_2)u_R'(\rho_2)}{\mathcal W}
u_L(\rho),
& \rho_1<\rho<\rho_2,
\\[1.4em]
\displaystyle
q_1
\left[
1+
\frac{a(\rho_1)u_L'(\rho_1)}{\mathcal W}
u_R(\rho)
\right]
+
q_2
\left[
1+
\frac{a(\rho_2)u_L'(\rho_2)}{\mathcal W}
u_R(\rho)
\right],
& \rho>\rho_2.
\end{cases}
\label{eq:global_two_charge_solution}
\end{align}
This solution obeys \eqref{eq:global_E_vanishing_BC}, as required by the vanishing of the boundary electric fields.

\paragraph{The on-shell energy.}
We evaluate the Hamiltonian on-shell:
\begin{align}
    H[\phi_0;q_1,q_2]
    &=
    \frac12
    \int d\rho\,a(\rho)
    \left[
        (\partial_\rho\phi_0)^2
        +
        \mu^2
        \left(\phi_0+\sqrt{\pi}\theta\right)^2
    \right]
    \nonumber\\
    &=\frac{\pi}{2\mathcal W}
    \Big[
        q_1^2a(\rho_1)^2u_L'(\rho_1)u_R'(\rho_1)
        +
        q_2^2a(\rho_2)^2u_L'(\rho_2)u_R'(\rho_2)
        \nonumber\\
        &\hspace{10em}
        +
        2q_1q_2a(\rho_1)a(\rho_2)
        u_L'(\rho_1)u_R'(\rho_2) \Big].
    \label{eq:global_two_charge_energy}
\end{align}
There are two special cases we consider. One has $u_L'(\rho)>0$, $u_R'(\rho)<0$, and $\mathcal W<0$, so the self energies are positive and the binding energy is indeed negative when $q_1q_2<0$.
\begin{enumerate}
    \item \textbf{$(q_1,q_2)=(q,-q)$: the neutral pair.}
    For a pair of equal and opposite charges,
    \begin{align}
        H[\phi_0;q,-q]
        =
        \frac{\pi q^2}{2\mathcal W}
        \Big[
            a(\rho_1)^2u_L'(\rho_1)u_R'(\rho_1)
            &+
            a(\rho_2)^2u_L'(\rho_2)u_R'(\rho_2)
            \nonumber\\&-
            2a(\rho_1)a(\rho_2)u_L'(\rho_1)u_R'(\rho_2)
        \Big].
    \end{align}

    \item \textbf{$(q_1,q_2)=(q,0)$: the single-charge self-energy.}
    For a single external charge at $\rho_0$,
    \begin{align}
        H[\phi_0;q,0]
        =
        \frac{\pi q^2}{2\mathcal W}
        a(\rho_0)^2u_L'(\rho_0)u_R'(\rho_0).
        \label{eq:global_self_energy}
    \end{align}
    Since the expression is quadratic in the charge, the same formula holds for
    a charge $-q$.
\end{enumerate}

\paragraph{The $q\bar q$ potential.}
The $q\bar q$ potential is again obtained from the total energy of a neutral pair by subtracting the one-body self-energies:
\begin{align}
    V_{q\bar q}(\rho_1,\rho_2)=
    H[\phi_0;q,-q]-H[\phi_0;q,0]_{\rho_1}-H[\phi_0;0,-q]_{\rho_2}.
\end{align}
Using \eqref{eq:global_two_charge_energy} and
\eqref{eq:global_self_energy}, this gives
\begin{align}
    V_{q\bar q}(\rho_1,\rho_2)
    =
    -\frac{\pi q^2}{\mathcal W}
    a(\rho_1)a(\rho_2)
    u_L'(\rho_1)u_R'(\rho_2).
    \label{eq:global_qbarq_potential}
\end{align}
For opposite charges this contribution is negative, corresponding to an attractive binding energy.

\paragraph{The electric field profile.}
We now specialize to an equal and opposite pair of external charges $q_1=-q_2=q$. Using $E(\rho)=\frac{\phi_0(\rho)}{\sqrt{\pi}}+\theta(\rho)$, the electric field profile becomes
\begin{align}
E(\rho)
=
\begin{cases}
\displaystyle
-\frac{q\,u_L(\rho)}{\mathcal W}
\left[
    a(\rho_1)u_R'(\rho_1)
    -
    a(\rho_2)u_R'(\rho_2)
\right],
& \rho<\rho_1,
\\[1.4em]
\displaystyle
-\frac{q}{\mathcal W}
\left[
    a(\rho_1)u_L'(\rho_1)u_R(\rho)
    -
    a(\rho_2)u_R'(\rho_2)u_L(\rho)
\right],
& \rho_1<\rho<\rho_2,
\\[1.4em]
\displaystyle
-\frac{q\,u_R(\rho)}{\mathcal W}
\left[
    a(\rho_1)u_L'(\rho_1)
    -
    a(\rho_2)u_L'(\rho_2)
\right],
& \rho>\rho_2 .
\end{cases}
\label{eq:global_flux_profile_exact}
\end{align}
This exact profile satisfies $E(\rho_1^+)-E(\rho_1^-)=q$ and $E(\rho_2^+)-E(\rho_2^-)=-q$ as well as $E(-\infty)=E(+\infty)=0$.

\paragraph{Comparison with the Schwarzschild frame.}
On a constant global time slice, the radial geodesic distance is
\begin{align}
    D(\rho_a,\rho_b)=L|\rho_b-\rho_a|,
    \qquad
    d=D(\rho_1,\rho_2).
\end{align}
In the Schwarzschild frame, the equation of motion \eqref{eq:EOM-Sch} in the coordinates $x=L\log r$ is manifestly invariant under shifts of $x$, as evident from:
\begin{align}
    \partial_r(f\partial_r\phi) = \partial_x^2\phi + \frac1L \partial_x\phi.
\end{align}
So, the Green's function depends only on differences of proper distance $d =x_2 - x_1 = L\log(\frac{r_2}{r_1})$ and the electric field profile organizes itself into a function of $d$. In contrast, the global frame differential operator $-\partial_\rho\left(\cosh\rho\,\partial_\rho\right) + \mu^2\cosh\rho$ is not invariant under translations in $\rho$. Consequently, the profile \eqref{eq:global_flux_profile_exact} depends not only on coordinate differences but also on the absolute positions of the charges through $u_L$, $u_R$, and their derivatives evaluated at $\rho_1$ and $\rho_2$. Near the two asymptotic boundaries, the modes reduce to pure exponentials:
\begin{align}
    u_R(\rho) &\underset{\rho\to+\infty}{\sim} e^{-(\nu+1)\rho},
    &
    u_L(\rho) &\underset{\rho\to-\infty}{\sim} e^{+(\nu+1)\rho}.
\end{align}
Thus, the exterior tails decay exponentially in the proper distance
\begin{align}
    E(\rho)
    &\propto
    \exp\left[
        -(\nu+1)\frac{D(\rho,\rho_2)}{L}
    \right],
    && \rho>\rho_2\gg1,
    \\
    E(\rho)
    &\propto
    \exp\left[
        -(\nu+1)\frac{D(\rho,\rho_1)}{L}
    \right],
    && \rho<\rho_1\ll-1.
\end{align}

The global and Schwarzschild calculations therefore agree on the exponential screening controlled by the massive bosonized mode, while differing in the exact spatial profile because the global redshift factor breaks translation invariance.

\subsection{Screening in QED$_2$ on an AdS$_2$ black hole: finite temperature}
In this subsection we show that the massless Schwinger model remains screened in an AdS$_2$ black-hole background, now diagnosed thermodynamically through the free energy of a static $q\bar q$ pair. That the conclusion is unchanged is to be expected, since the black hole simply restricts the theory to the Schwarzschild patch and endows it with a temperature, and we saw above that the massless model already screens in that patch at zero temperature. The value of the calculation is therefore methodological. It provides a finite-temperature, free-energy probe of confinement that remains well defined in the presence of a horizon and can be applied to models with richer screening behaviour, such as the massive theory or non-abelian generalizations. As with the zero-temperature energy \eqref{eq:TotalEnergy}, the free energy contains the self-energies of the individual probes; once these are subtracted, the remainder is the $q\bar q$ potential, which again tends to zero as the charges are separated.

A word is in order on what we mean by a ``black hole'' in two dimensions. Pure two-dimensional gravity is topological, since the Einstein-Hilbert action is the Euler characteristic, so the metric \eqref{eq:SchwarzschildBH-2} does not by itself solve any gravitational dynamics, and as a geometry it is merely a patch of AdS$_2$: it has constant negative curvature, no curvature singularity, and its ``horizon'' at $r=r_h$ is a bifurcate Killing horizon rather than the event horizon of a causally hidden interior. A genuine black-hole interpretation, with a horizon entropy and a notion of the interior, is supplied by coupling the metric to a dilaton, as in JT gravity \cite{Jackiw:1984je, Teitelboim:1983ux, Almheiri:2014cka, Maldacena:2016upp}; a Hamiltonian treatment in which the gravitational sector is integrated out by solving the constraints is developed in \cite{Kraus:2025}. In the absence of gravitational backreaction, none of this gravitational data changes the conclusions about the absence of confinement, which requires only a choice of preferred time. We take this to be the Hartle-Hawking state, prepared by the Euclidean path integral with the smooth cap that fixes the temperature to the Hawking value $T=r_h/2\pi L^2$.

\paragraph{Free energy in the Schwarzschild frame.}
We begin by computing the free-energy cost of inserting two static test charges in the Schwarzschild frame. In order to prepare the Hartle-Hawking vacuum, we continue to Euclidean time \(t=-i\tau_{E}\), for which the metric is
\begin{equation}
    ds_E^2
    =
    f(r)d\tau_{E}^2
    +
    \frac{1}{f(r)}dr^2,
\end{equation}
with \(\tau_{E}\sim\tau_{E}+\beta\). As we will see, for the Hartle-Hawking state, regularity at the Euclidean horizon fixes \(\beta\) to be the inverse Hawking temperature~\cite{HartleHawking1976,GibbonsHawking1977}. We work in axial gauge, where field configurations take the form 
\begin{equation}
    A=\varphi(r, \tau_E)\;d\tau_{E} .
\end{equation}
Here, $\varphi$ denotes the Euclidean space gauge field, and should not be confused with the $\phi$ from the preceding subsections. Following \cite{Gross:1995bp}, the quadratic Euclidean effective action for \(\varphi\) is
\begin{equation}
    S_E[\varphi]
    =
    \int_0^\beta d\tau_{E}
    \int_{r_h}^{\infty}dr
    \left[
        \frac{1}{2e^2}\left(\partial_r\varphi\right)^2
        +
        \frac{1}{2\pi}\frac{\varphi^2}{f(r)}
    \right].
\end{equation}
Since time is periodic $\tau_E\sim \tau_E+\beta$, we fourier expand as
\begin{align}
    \varphi(r, \tau_E) &= \sum_{n\in \mathbb Z}\varphi_n(r) e^{i\omega_n \tau_E}, & \omega_n &=\frac{2\pi}{\beta} n.
\end{align}
This yields
\begin{align}\label{eq:free_energy_wo_source}
    S_E[\varphi]
   & =
    \beta\sum_{n}\mathcal F_n[\varphi_n],& \mathcal F_n[\varphi_n]
    &=
    \int_{r_h}^{\infty}dr
    \left[
        \frac{1}{2e^2}|\partial_r\varphi_n|^2
        +
        \frac{1}{2\pi}\frac{|\varphi_n|^2}{f(r)}
    \right].
\end{align}
We now insert two static probe charges,
\begin{equation}
    \rho(r)
    =
    q\delta(r-r_1)-q\delta(r-r_2).
\end{equation}
In the Euclidean path integral, these charges are represented by Wilson lines
wrapping the thermal circle:
\begin{equation}
    W_\rho
    =
    \exp\left[
        i\int_0^\beta d\tau_{E}
        \int_{r_h}^{\infty}dr\,
        \rho(r)\varphi(r)
    \right] = \exp\left[
        i\beta
        \int_{r_h}^{\infty}dr\,
        \rho(r)\varphi_0(r)
    \right] \,,
\end{equation}
where only the zero mode survives the integration. The shift in the free energy due to inserting probes is obtained by comparing the thermal partition function with and without the static probe charges. With the probes inserted, the partition function is
\begin{align}
    Z_\rho
    =
    \int D\varphi\, e^{-S_E[\varphi]} W_\rho[\varphi].
\end{align}
Without the probes, the partition
function is \(Z_{\rho = 0}\). Therefore, the Wilson-line expectation value in the thermal ensemble without the probes is
\begin{align}
    \langle W_\rho\rangle
    =
    \frac{1}{Z_{\rho = 0}}
    \int D\varphi\, e^{-S_E[\varphi]} W_\rho[\varphi]
    =
    \frac{Z_\rho}{Z_{\rho = 0}}.
\end{align}
Since the free energy is \(\mathcal F=-\beta^{-1}\log Z\), the free-energy cost
of inserting the probes is
\begin{align}
    \Delta\mathcal F[\rho]
    =
    -\frac{1}{\beta}\log\frac{Z_\rho}{Z_{\rho = 0}}
    =
    -\frac{1}{\beta}\log\langle W_\rho\rangle .
\end{align}
Note that the ratio is determined purely by the functional integral over the zero mode, i.e.
\begin{align}
    \frac{Z_\rho}{Z_{\rho = 0}} = \frac{\int D\varphi_0\exp\{-\beta \mathcal F_0[\varphi_0]+i\beta\int_{r_h}^{\infty}dr\,\rho(r)\varphi_0(r)\}}{\int D\varphi_0 \exp\{-\beta \mathcal F_0[\varphi_0]\}}\,.
\end{align}
Equivalently, after integrating by parts and using the boundary conditions (to be justified shortly) that $\varphi(r_h) = \varphi(\infty) = 0$,
\begin{equation}
    \mathcal F_0[\varphi_0]
    =
    \frac{1}{2e^2}
    \int_{r_h}^{\infty}dr\,
    \varphi_0(r)
    \left[
        -\frac{d^2}{dr^2}
        +
        \frac{m_\gamma^2}{f(r)}
    \right]
    \varphi_0(r),
    \qquad
    m_\gamma^2\equiv\frac{e^2}{\pi}.
\end{equation}
Using the quadratic effective action for \(\varphi_0\), the Wilson-line
expectation value is Gaussian:
\begin{align}
    \langle W_\rho\rangle
    =
    \frac{
    \int D\varphi_0\,
    \exp\left[
        -\beta \mathcal F_0[\varphi_0]+i\beta\int_{r_h}^{\infty}dr\,\rho(r)\varphi_0(r)
    \right]
    }{
    \int D\varphi_0\,
    \exp\left[
        -\beta \mathcal F_0[\varphi_0]
    \right]
    } .
\end{align}
Performing the Gaussian integral gives
\begin{align}
    \log \langle W_\rho\rangle
    =
    -\frac{\beta e^2}{2}
    \int_{r_h}^{\infty}dr
    \int_{r_h}^{\infty}dr'\,
    \rho(r)G_\beta(r,r')\rho(r'),
\end{align}
where \(G_\beta(r,r')\) is the Green's function of the static operator
\begin{align}\label{eq:Free_Energy_Saddle}
    \left[
        -\frac{d^2}{dr^2}
        +
        \frac{m_\gamma^2}{f(r)}
    \right]G_\beta(r,r')
    =
    \delta(r-r').
\end{align}
Therefore,
\begin{align}
    \Delta\mathcal F[\rho]
    =
    -\frac{1}{\beta}\log\langle W_\rho\rangle
    =
    \frac{e^2}{2}
    \int_{r_h}^{\infty}dr
    \int_{r_h}^{\infty}dr'\,
    \rho(r)G_\beta(r,r')\rho(r').
\end{align}
The Green's function is dependent on the boundary conditions at the black-hole horizon and the AdS$_2$ boundary.

\paragraph{Boundary Conditions.}
The boundary condition on the black hole horizon is fixed by the choice of vacuum. Choosing the Hartle-Hawking state is equivalent to choosing a periodicity for the Euclidean time which removes any conical singularity in the metric. Explicitly, if we expand the Euclidean metric near the black hole horizon (introducing $y=r-r_h$) the line element becomes

\begin{equation}
    ds^2 = f(r(y))d\tau_{E}^2+\frac{1}{f(r(y))}dy^2~.
\end{equation}
We note that
\begin{equation}
    f(r(y))=\frac{r^2-r_h^2}{L^2}=\frac{y(r+r_h)}{L^2}=y\frac{2r_h}{L^2}+{O}(y^2) = yf'(r_h) + O(y^2).
\end{equation}
Next, we introduce $R$ via $y=\frac{1}{4}f'(r_h)R^2$. The metric becomes
\begin{equation}
    ds^2 =  dR^2+R^2\left(\frac{f'(r_h)}{2}d\tau_{E}\right)^2 + \dots,
\end{equation}
implying that the conical singularity disappears if $\frac{f'(r_h)}{2}\tau_{E}$ has periodicity $2\pi$, or equivalently that $\tau_{E}$ has periodicity $\frac{4\pi}{f'(r_h)}=\frac{2\pi L^2}{r_h}$. Thus, by choosing the Hartle-Hawking vacuum (equivalently a temperature of $T=\frac{r_h}{2\pi L^2}$) we can reasonably impose the requirement that the one-form $A$ is not singular at the horizon. This is only possible if $\varphi(r_h)=0$. 

\begin{figure}
    \centering
    \includegraphics[width=0.6\linewidth]{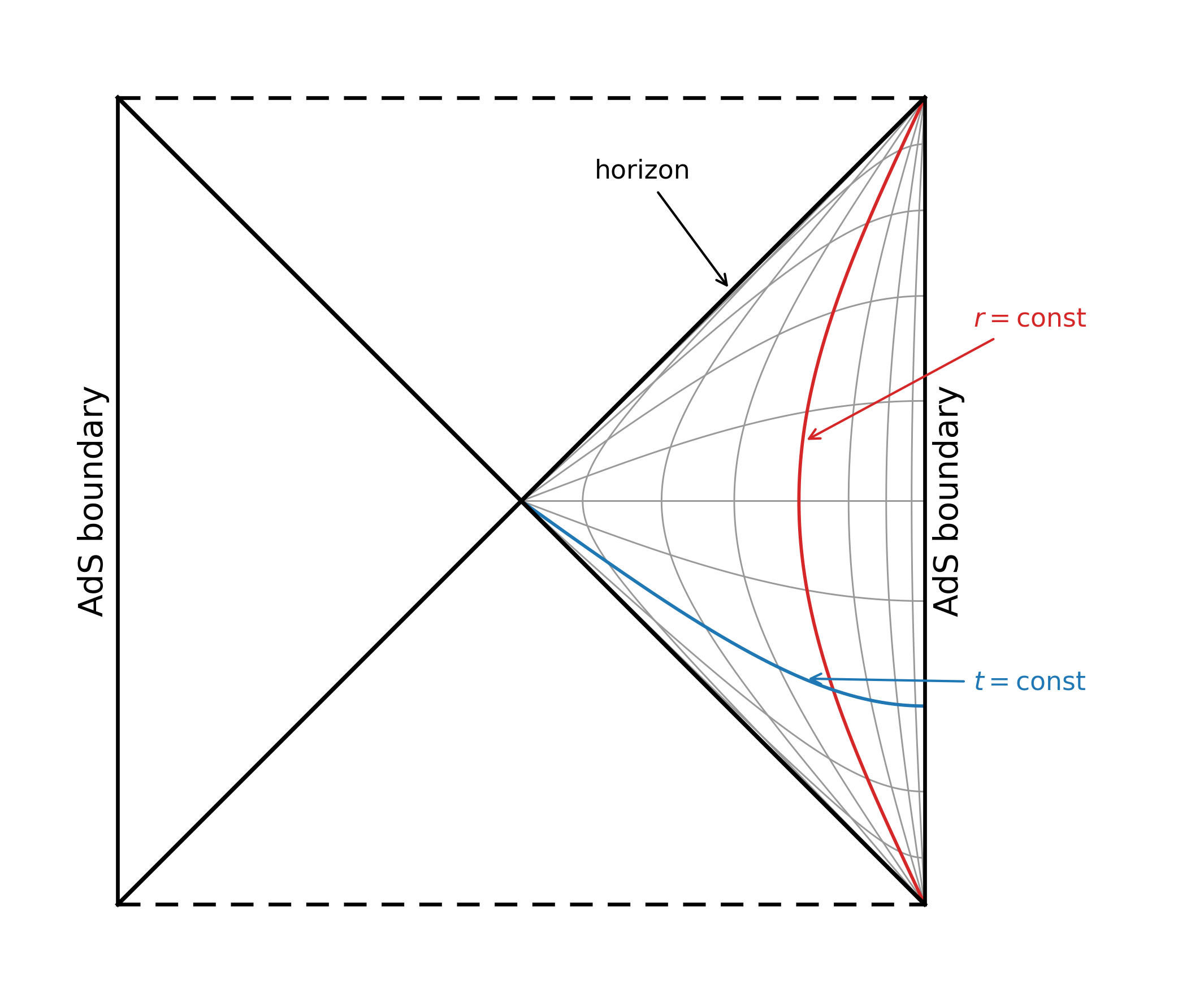}
    \caption{Penrose diagram of AdS$_2$ Schwarzschild black hole. Timelike curves are $r=\text{constant}$ (e.g. the red curve) and spacelike curves are $t=\text{constant}$ (e.g. the blue curve).}
    \label{fig:penrose}
\end{figure}

The AdS$_2$ boundary is more subtle (see figure \ref{fig:penrose}). We can look at asymptotic homogeneous solutions to \eqref{eq:Free_Energy_Saddle}. In that limit the equation becomes

\begin{equation}
        -u''(r)+\frac{m_{\gamma}^2 L^2}{r^2}u(r)=0~.
\end{equation}
Making the ansatz $u(r)=r^{\lambda}$ we see that the general solution to this equation is
\begin{equation}
    u(r)=c_1r^{\frac{1}{2}+\Delta}+c_2 r^{\frac{1}{2}-\Delta},\qquad\Delta=\sqrt{\frac{1}{4}+m_{\gamma}^2L^2}~.
\end{equation}
In order for our free energy to be finite, we require that the potential does not become infinite at the AdS$_2$ horizon. Hence we impose that $c_1=0$. We will thus choose a Green's function which is zero at the black hole horizon, and normalizable at the AdS$_2$ horizon. 
\paragraph{The static potential at finite temperature.}
We now evaluate the fixed-charge free energy using the static Green's function
with regularity at the Euclidean black-hole horizon and normalizability at the AdS$_2$ boundary. Two convenient homogeneous solutions are
\begin{align}
    u_H(r)
    &=
    \sqrt{x^2-1}\,P_\nu^1(x),
    &
    u_\infty(r)
    &=
    -\sqrt{x^2-1}\,Q_\nu^1(x),
    \qquad
    x=\frac{r}{r_h},
\end{align}
where
\begin{align}
    \nu=\Delta-\frac12,
    \qquad
    \Delta=\sqrt{\frac14+m_\gamma^2L^2}.
\end{align}
The solution \(u_H\) satisfies the horizon boundary condition, while
\(u_\infty\) satisfies the AdS-boundary condition. Defining the Wronskian
\begin{align}
    \mathcal W
    =
    u_H'(r)u_\infty(r)
    -
    u_H(r)u_\infty'(r),
\end{align}
the Green's function is
\begin{align}
    G_\beta(r,r')
    =
    \frac{u_H(r_<)u_\infty(r_>)}{\mathcal W},
    \qquad
    r_<\equiv \min(r,r'),
    \qquad
    r_>\equiv \max(r,r').
\end{align}
For a neutral pair of static probe charges,
\begin{align}
    \rho(r)=q\delta(r-r_1)-q\delta(r-r_2),
\end{align}
the Wilson-line free-energy shift is
\begin{align}
    \Delta\mathcal F(r_1,r_2)
    =
    \frac{e^2}{2}
    \int_{r_h}^{\infty}dr
    \int_{r_h}^{\infty}dr'\,
    \rho(r)G_\beta(r,r')\rho(r').
\end{align}
Substituting the explicit charge density gives
\begin{align}
    \Delta\mathcal F(r_1,r_2)
    =
    \frac{e^2q^2}{2}
    \left[
        G_\beta(r_1,r_1)
        +
        G_\beta(r_2,r_2)
        -
        2G_\beta(r_1,r_2)
    \right].
\end{align}
This is the total free-energy cost of inserting the pair and includes the
self-energies of the individual probe charges. The one-body self-energy of a
probe charge at \(r_0\) is
\begin{align}
    \Delta\mathcal F_{\rm self}(r_0)
    =
    \frac{e^2q^2}{2}G_\beta(r_0,r_0).
\end{align}
Therefore, the self-energy-subtracted \(q\bar q\) potential is identified at finite temperature as:
\begin{align}
    V_{q\bar q}(r_1,r_2)
    &=
    \Delta\mathcal F(r_1,r_2)
    -
    \Delta\mathcal F_{\rm self}(r_1)
    -
    \Delta\mathcal F_{\rm self}(r_2)=
    -e^2q^2G_\beta(r_1,r_2).
\end{align}
The finite-temperature binding energy is therefore controlled by the
off-diagonal Green's function connecting the two probe locations.

\paragraph{Remarks on the temperature dependence.} The temperature enters only through the horizon radius,
\begin{align}
    T=\frac{r_h}{2\pi L^2},
\end{align}
and therefore only through the Green's function \(G_\beta(r,r')\). This yields
\begin{align}
    V_{q\bar q}(r_1,r_2)=-e^2q^2G_\beta(r_1,r_2),
\end{align}
In the zero-temperature limit \(r_h\to0\), the
Green's function reduces to the pure AdS\(_2\) Schwarzschild-frame Green's
function, giving
\begin{align}
    V_{q\bar q}(r_1,r_2)
    \longrightarrow
    -\frac{e^2q^2}{2\nu+1}\,
    r_1\left(\frac{r_1}{r_2}\right)^\nu,
    \qquad r_1<r_2.
\end{align}
At finite temperature, including the large-temperature regime, the same
subtracted expression remains valid, with \(G_\beta\) replaced by the Green's
function satisfying the corresponding horizon and AdS-boundary conditions.
Thus temperature changes the quantitative profile of the interaction, but not
the conclusion: after subtracting the probe self-energies, the massless theory
remains screened rather than confining.

\subsection{Consistency with the flat-space Schwinger Model}\label{sec:flatspace}
Having obtained the static potential in the Schwarzschild and global frames, we close the section by checking that both reduce to the known flat-space massless Schwinger model in the appropriate limit. In each case, the flat-space theory is recovered by zooming into a local patch and sending the AdS$_2$ radius $L\to\infty$ at fixed geodesic separation $d$, so that the curvature scale decouples while the dynamically generated Schwinger mass $m_\gamma=e/\sqrt{\pi}$ is held fixed. We wish to reproduce the static potential of the flat-space massless Schwinger model~\cite{Gross:1995bp}
\begin{align}\label{eq:V_flat}
    V_{\rm flat}(d)
    = \frac{\sqrt{\pi}}{2}\,e q^2\left(1 - e^{-m_\gamma d}\right),
\end{align}
a Yukawa-screened interaction whose binding piece $-\tfrac{\sqrt{\pi}}{2}e q^2\,e^{-m_\gamma d}$ decays exponentially and saturates to the constant $\tfrac{\sqrt{\pi}}{2}e q^2$ at large separation, the hallmark of a screened phase. In the flat space limit, we find that the self-energies collapse to  position-independent constants, as required by flat-space translational symmetry.

\paragraph{The Schwarzschild AdS$_2$ frame.}
To recover the flat-space Schwinger model we zoom in on a local patch by writing $r = L + x$ and sending $L\to\infty$ with the geodesic separation $d = L\log(r_2/r_1)$ held fixed. Writing the dynamically generated mass as $m_\gamma \equiv e/\sqrt{\pi}$, $\nu$ grows without bound,
\begin{align}
    \nu = -\tfrac12 + \sqrt{\tfrac14 + m_\gamma^2 L^2}
    \;\longrightarrow\; m_\gamma L 
\end{align}
while the coordinate ratio exponentiates into the proper distance,
\begin{align}
    \left(\frac{r_1}{r_2}\right)^\nu
    = \exp\!\left[-\frac{\nu d}{L}\right]
    \;\longrightarrow\; e^{-m_\gamma d}.
\end{align}
The vacuum energy of the neutral pair as $L\to\infty$ is
\begin{align}
    H[\phi_0;q,-q]
    = \frac{e^2 q^2}{2(2\nu+1)}
      \left[\,r_2 + r_1\!\left(1 - 2\left(\frac{r_1}{r_2}\right)^\nu\right)\right]
    \;\longrightarrow\;
    \frac{\sqrt{\pi}\,e\,q^2}{2}\left(1 - e^{-m_\gamma d}\right),
\end{align}
 which is precisely the flat-space result~\eqref{eq:V_flat}: a Yukawa-screened interaction that saturates to the sum of the self energies $\tfrac{\sqrt{\pi}}{2} g q^2$ at large separation. Unlike in AdS$_2$, the self energies are position independent constants in flat space due to translational invariance.

\paragraph{The global AdS$_2$ frame.}
The flat-space Schwinger model is recovered as in the Schwarzschild frame, with one instructive difference of normalization. Zooming into a local patch by writing $\rho=\rho_*+x/L$ and sending $L\to\infty$, the leading order metric becomes:
\begin{align}
    ds^2 &= -dt_{\rm Minkowski}^2 + dx^2  + O\left(\frac x L\right) ,&t_{\rm Minkowski}&= L\cosh\rho_*\,\tau.
\end{align}
Setting $\rho_i = \rho_* + x_i/L$ with the geodesic separation $d=L|\rho_2-\rho_1|$ held fixed, the static potential in the flat-space
\begin{align}
    E_{\rm vac}^{\rm Minkowski} = \lim_{L\rightarrow\infty }\frac{H[\phi_0; q,-q]}{L\cosh\rho_*} = \frac{\sqrt{\pi}}{2}eq^2\left(1- e^{-m_\gamma d}\right).
\end{align}
confirming that the two frames agree on the energy.

\section{Tensor-Network Simulations of String Breaking: Static Potential and \\ Real-Time Dynamics}
\label{sec:Simulation}

In this section, we validate the theoretical findings presented in Sec.~\ref{sec:Screening} and further investigate the massive model for which analytical results only exist in the small-mass expansion \cite{Alimohammadi:2000fg}. First, we analyze the scaling of the static potential as a function of the geodesic separation between the charges, illustrating explicitly that while the vacuum energy may vanish or diverge depending on the specific details of how the geodesic separation is taken to infinity, the binding potential itself is always screened, demonstrating that the theory is screened (Sec.~\ref{sec:StaticSimulation}). Furthermore, we simulate the profile of the putative electric flux tube, which, at strong-coupling, develops a node, contrary to what is expected of a genuinely confined electric flux tube (Sec.~\ref{sec:flux_simulation}). This further confirms the screened phase of the theory. To ensure consistency with theoretical results, we match our results to the continuum electric field profile. Lastly, we present a real-time simulation of string breaking (Sec.~\ref{sec:real_time}).

All numerical results presented in this section were obtained via a tensor-network approach using a Matrix Product State (MPS) ansatz optimized by the Density Matrix Renormalization Group (DMRG) algorithm.
\begin{figure}[b]
    \centering
    \includegraphics[width=0.5\linewidth]{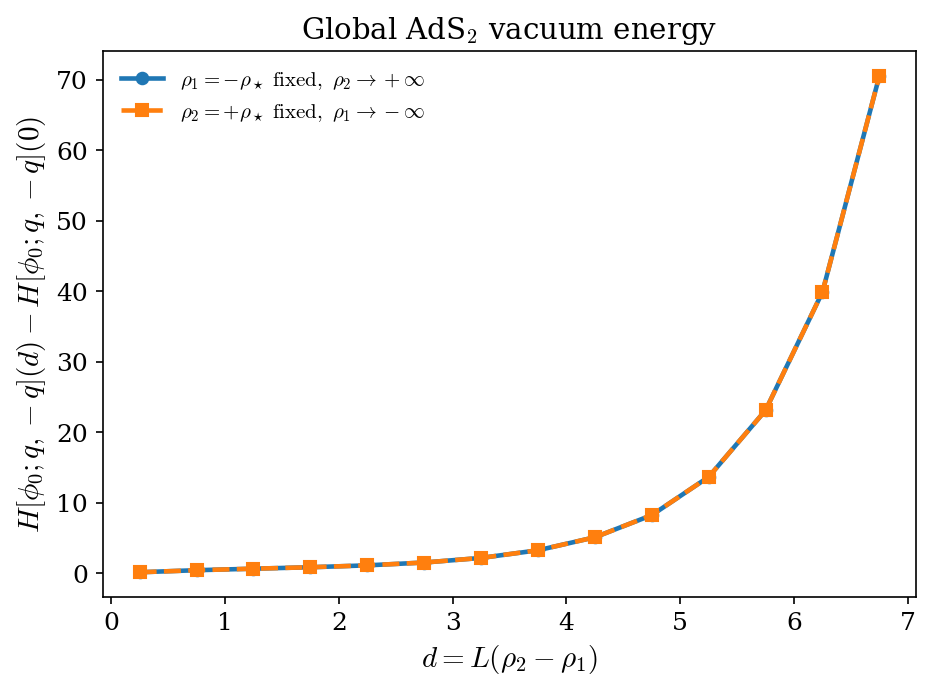}\includegraphics[width=0.5\linewidth]{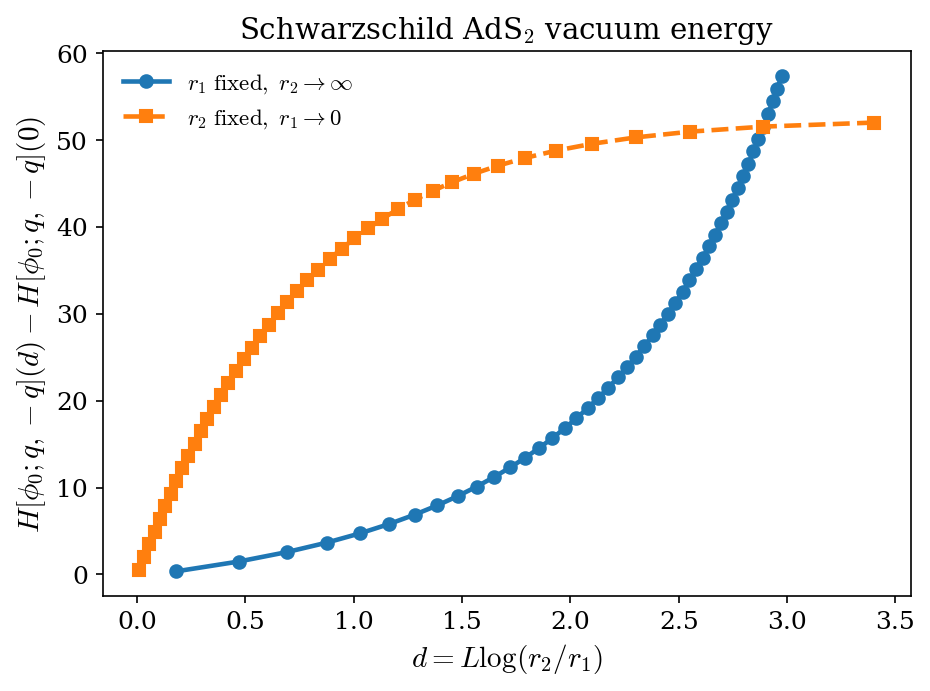}
    \caption{The above figure shows plots of the vacuum energy in the presence of a $q\bar q$-pair as a function of their geodesic separation in both global AdS$_2$ (left) and Schwarzschild (right) frames \textit{prior} to the self-energy subtraction.}
    \label{fig:VacuumEnergy}
\end{figure}
\subsection{Static Potential in AdS$_2$}\label{sec:StaticSimulation}

In this subsection, we analyze the vacuum energy and static potential of QED$_2$ on AdS$_2$ in the presence of a pair of equal and opposite static charges. First, we analyze the vacuum energy of this configuration, which, in flat space, coincides with the static potential up to translation-invariant (hence, constant) self-energies. On the other hand, as we established theoretically in the preceding section, since the na\"ive translational invariance is broken in the presence of a curved background, these contributions to the vacuum energy are no longer constant. If one does not carefully subtract these contributions, one would incorrectly conclude that the static potential is diverging (confined) for some cases, whilst being finite (screened) in other cases. Moreover, this is also in apparent conflict with the explicitly broken one-form symmetry; this is expected to screen any integer-charge electric flux tubes.

\begin{figure}[h]
    \centering    \includegraphics[width=0.5\linewidth]{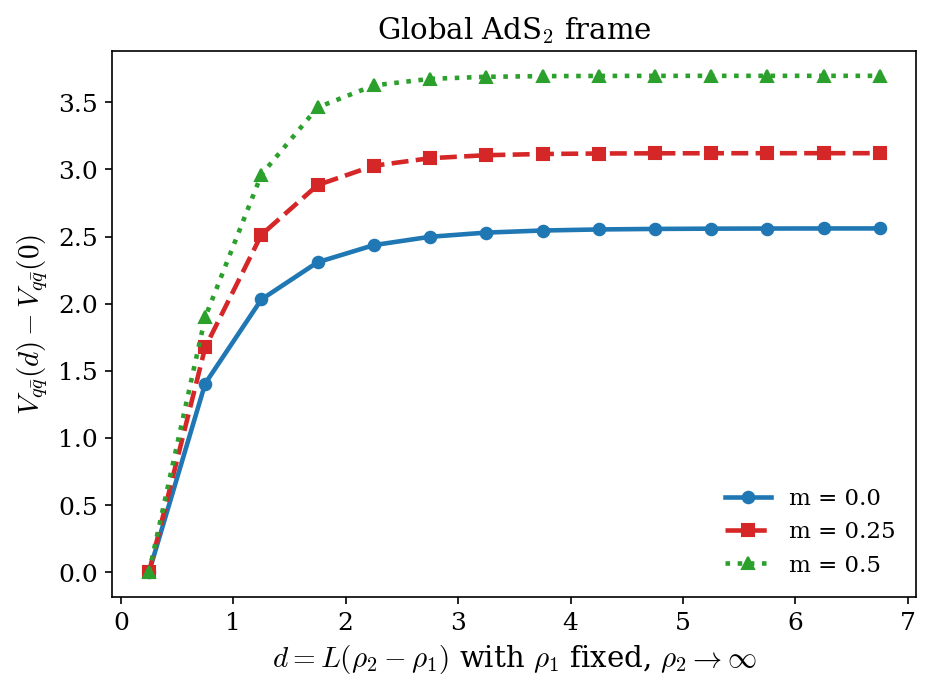}\includegraphics[width=0.5\linewidth]{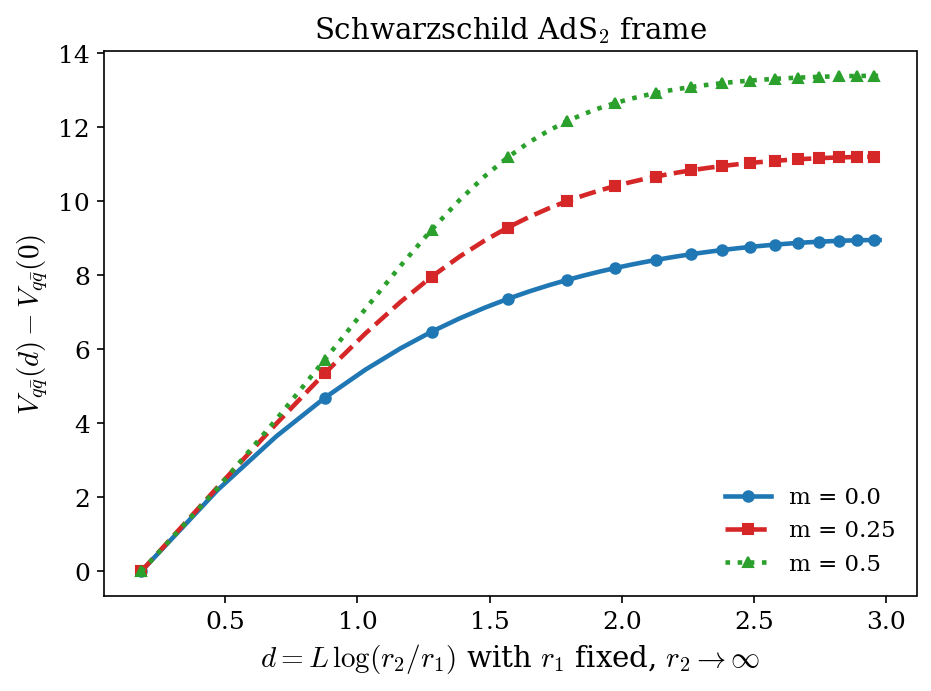}
    \caption{Static $q\bar q$ potential $V_{q\bar q}(d)-V_{q\bar q}(0)$, obtained from the ground-state energy of a static charge pair by subtracting the (screened) single-charge self-energies, for masses $m=0,\,0.25,\,0.5$. Left: global AdS$_2$ with fixed $\rho_1$ ($N=48$, $a=0.25$, geodesic separation $D=L(\rho_2-\rho_1)$). Right: Schwarzschild AdS$_2$ frame with fixed $r_1$ ($N=100$, $a=1$, separation $d=L\ln(r_2/r_1)$). Lattice spacing $a$ in the respective radial coordinate; $L=1$, $e=2.51$ throughout. }
    \label{fig:screening}
\end{figure}

Figure~\ref{fig:VacuumEnergy} illustrates the apparent tension between the global AdS$_2$ and Schwarzschild frames. Na\"ively expecting this vacuum energy to reproduce the static binding potential yields results which are in apparent contradiction. The global AdS$_2$ potential is diverging as a function of the geodesic separation $d$ between the probe charges. On the other hand, the Schwarzschild frame potential may be diverging or finite depending on precisely how we approach $d=\infty$. When Schwarzschild radial position $r_1$ is held fixed and $r_2\rightarrow\infty$ ($d\rightarrow\infty$), the vacuum energy diverges, and the theory is \textit{seemingly} confined. On the other hand, if one fixes $r_2$ and sends $r_1\rightarrow0$ (also $d\rightarrow\infty$), it is finite, and the theory is \textit{seemingly} screened.

As argued in Sec.~\ref{sec:Screening}, the resolution is to carefully account for the self-energy contributions, which are non-constant in curved space. Figure~\ref{fig:screening} shows the static potential after the self-energy subtraction. The left panel considers global AdS$_2$ with $\rho_1$ fixed and $\rho_2\rightarrow\infty$, which one would na\"ively have concluded to be confining. However, upon suitable subtraction of self-energy contributions, we find it to be screening, in agreement with theoretical predictions. The right panel considers the same scenario in the Schwarzschild frame with $r_1$ fixed and $r_2\rightarrow\infty$: indeed, this also turns out to be screened.

\subsection{Electric flux profile and agreement with the continuum limit}
\label{sec:flux_simulation}
We now the simulate the electric field and match it against the continuum prediction.
\begin{figure}[ht]
    \centering
    \includegraphics[width=0.47\linewidth]{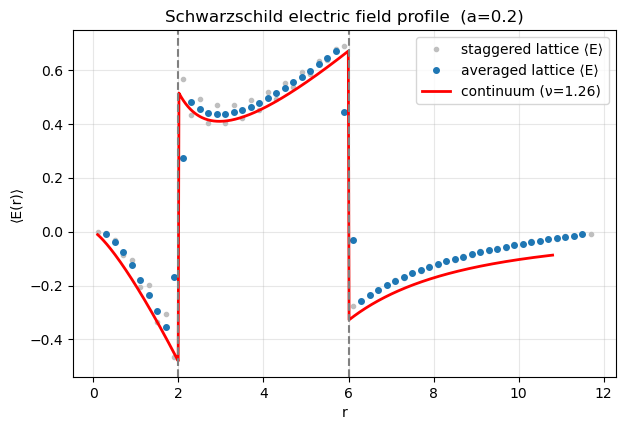}\hfill
    \includegraphics[width=0.5\linewidth]{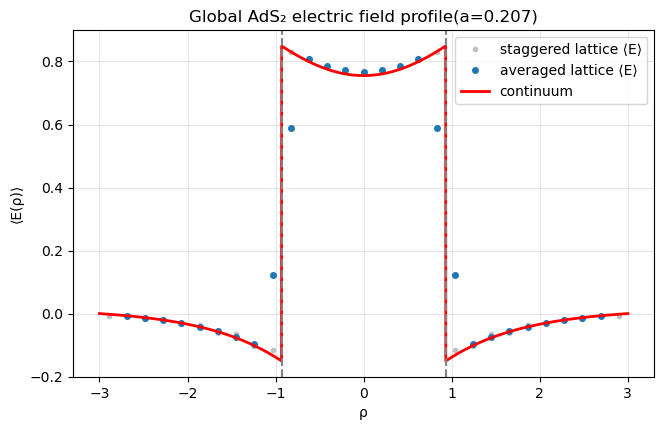}
    \caption{Electric flux-tube profile of a neutral probe pair $q_1=-q_2=q$, comparing the
    DMRG lattice simulation against the continuum bosonization prediction.
    {Left:} Schwarzschild frame ($a=0.2$), with charges at $r_1=2$, $r_2=6$ (dashed) and
    $\nu\simeq1.26$. {Right:} global AdS$_2$ frame ($a=0.207$), with charges at
    $\rho_1=-1$, $\rho_2=+1$ (dashed). Faint gray points are the raw staggered $\langle E\rangle$,
    blue points are the binomial-averaged (de-staggered) field, and the red curve is the
    continuum profile with $\nu$ fixed by the coupling. The de-staggered data track the continuum
    curve in all regions with no free normalization. In both frames the field changes sign across
    the insertions and decays exponentially in the exterior, while the symmetry of the global
    profile, in contrast to the geodesic-distance organization of the Schwarzschild profile,
    reflects the $\cosh\rho$ redshift factor and the two asymptotic boundaries of the global slice.}
    \label{fig:flux-matching}
\end{figure}
Figure~\ref{fig:flux-matching} shows the vacuum expectation value $\langle E\rangle$ for an equal-and-opposite pair $q_1=-q_2=q$ in the Schwarzschild frame (left) and the global AdS$_2$ frame (right). In each panel, the faint gray points are the expectation values, which appear in a sawtooth pattern due to the use of staggered fermions. The blue points are obtained by de-staggering the electric field via:
\begin{equation}
    \bar E_n \;=\; \tfrac14\left(E_{n-1} + 2E_n + E_{n+1}\right),
    \label{eq:destagger}
\end{equation}
The solid red curve is the continuum profile of Sec.~\ref{sec:Screening}. After de-staggering, the lattice data fall onto the continuum curve, with expected deviations where the electric field encounters a discontinuity due to the probe charges $E(r_i^+)-E(r_i^-)=\pm q$.  As elucidated analytically in Fig.~\ref{fig:flux-tube-profile}, the profile is not the spatially constant field of a confining string. Between the two charges, the electric field decays exponentially as the coupling $e^2$ is raised. Outside the two charges, the electric field decays exponentially in the geodesic distance to the nearest insertion, indicating a screening cloud generated by the dynamical fermions. Both the Schwarzschild and global vacua yield different electric field values, but they agree that the theory is screening. 

\subsection{Real-time dynamics of string breaking}\label{sec:real_time}

We now present our results on a real-time simulation of a string breaking process. The initial state is constructed of the global AdS$_2$ vacuum $\ket{\Omega_{\rm Global}}$ as follows:
\begin{align}
    \ket{\Psi(0)} &= \chi_p^\dagger \mathcal{W}_{p, q} \chi_q\ket{\Omega_{\rm Global}} & \mathcal{W}_{p, q}&=\prod_{k=p}^{q-1}U_k
\end{align}
\begin{figure}[b]
    \centering
    \includegraphics[width=\linewidth]{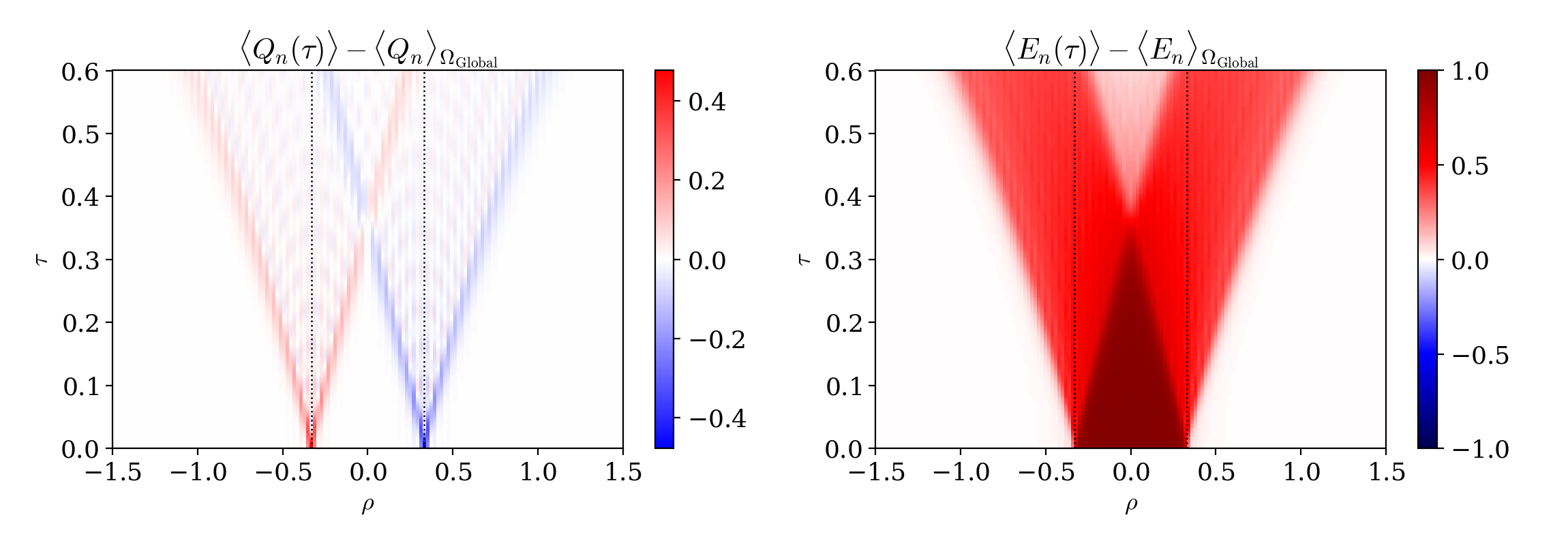}
    \caption{Real-time screening of an $e^{+}e^{-}$ pair of dynamical fermions in global AdS$_2$ by pair production out of the vacuum, in agreement with theoretical predictions. Left: the vacuum subtracted charge density $\langle Q_n(\tau)\rangle - \langle Q_n\rangle_{\Omega_{\rm Global}}$; right: the vacuum-subtracted electric field $\langle E_n(\tau)\rangle - \langle E_n\rangle_{\Omega_{\rm Global}}$, both as functions of the global radial coordinate $\rho$ and time $\tau$ for the massive Schwinger model on global AdS$_2$ discretized with $N=300$ staggered sites ($e=2$, $m=0.5$, $L=1$, $\rho_{\max}=3$). At $\tau=0$ a static pair is inserted at $\rho=\mp\tfrac13$ (dotted lines) joined by an electric flux string $\langle E_n\rangle\simeq1$ (right panel, dark core). Under two-site TDVP evolution ($d\tau=1.2\times10^{-3}$, bond dimension $\chi\le400$) the string breaks: charges produced from the vacuum screen the sources, the interior flux tube shrinks to zero, and the charge fronts propagate outward along the light cone. The per-site charge density peaks near $0.48$ because each unit charge is spread over a few lattice sites.}
    \label{fig:string-breaking}
\end{figure}
This initial state represents a gauge-invariant meson: an electron-positron pair that is initially bound together by a Wilson line, i.e. the electric string. Such a string is expected to be unstable, and hence decays into other strings as time evolves. To see this in a simulation, we compute:
\begin{align}
    \langle Q_n(\tau)\rangle  &= \bra{\Psi(\tau)}Q_n\ket{\Psi(\tau)}, & \langle E_n(\tau)\rangle  &= \bra{\Psi(\tau)}E_n\ket{\Psi(\tau)}, & \ket{\Psi(\tau)} &= e^{-i H\tau}\ket{\Psi(0)}.
\end{align}
where $Q_n$ refers to the staggered charge density. To see the string breaking process clearly without the expected staggering effects, we plot the vacuum-subtracted observables:
\begin{align}
    &\langle Q_n(\tau)\rangle - \langle Q_n\rangle_{\Omega_{\rm Global}} &  &\langle E_n(\tau)\rangle - \langle E_n\rangle_{\Omega_{\rm Global}}
\end{align}

In Fig.~\ref{fig:string-breaking}, we present a real-time simulation of the evolution of the electric charge and field densities, respectively. The meson is released into the interacting vacuum of the massive Schwinger model on global AdS$_2$. At $\tau=0$ the two charges sit at $\rho=\mp\tfrac{1}{3}$, bound by a uniform electric flux tube with $\langle E_n\rangle\simeq 1$ threaded between them by Gauss's law, while the charge density $\langle Q_n\rangle$ is sharply localized at the endpoints and integrates to $\pm 1$ on either side of the midpoint. As the state evolves, the string breaks: the field energy stored between the sources produces dynamical $e^{+}e^{-}$ pairs, and the newly produced charges rearrange so as to screen the original pair. This is visible as the steady decrease of the electric field in the region between the initial locations of the two charges. This is most pronounced beyond $\tau\simeq 0.35$, where the central field collapses toward zero. The interior screens rather than remaining confined, which is exactly what one expects since the fermions are massive and dynamical. This furnishes a fully nonperturbative, real-time demonstration of Schwinger pair production and string breaking on a curved AdS$_2$ background.

\section{Conclusion}
In this work, we analyzed confinement and screening in single-flavor QED$_2$ on AdS$_2$, with and without a Schwarzschild black hole, in the continuum and on the lattice. Our analysis was organized around two frames adapted to distinct choices of preferred time: the Schwarzschild and the global AdS$_2$ frames~\cite{Spradlin:1999bt}. This allowed us to disentangle which features of the static potential are physical and which are artifacts of the frame and its associated notion of energy.

On the analytic side, we settled the long-standing confinement-versus-screening puzzle for the massless theory on AdS$_2$. Earlier treatments identified the ground-state energy in the presence of the probe charges directly with the static potential and, on this basis, reached frame- and limit-dependent conclusions that appeared mutually inconsistent~\cite{Alimohammadi:2000fg, MohseniSadjadi:2000mg, MohseniSadjadi:2000nt, Gass:AdS2, Ghosh:1996}, some pointing to confinement and others to screening even as the geodesic separation was sent to infinity. We showed that the resolution is that, unlike in flat space, the probe self-energies in curved space are position dependent and must be subtracted before the binding potential can be read off. Solving the bosonized semiclassical theory~\cite{Coleman:1975pw, Coleman:1976uz} exactly in both frames, we found that once these self-energies are removed the binding potential saturates and the theory is screened, independently of how the geodesic separation is taken to infinity. This is compatible with the explicit breaking of the $\mathrm{U}(1)$ electric one-form symmetry by the dynamical fermions~\cite{Gaiotto:2014kfa, Gross:1995bp}, and it removes the apparent tension between the two frames. We obtained closed-form expressions for both the static $q\bar q$ potential and the electric flux-tube profile, and, by preparing the Hartle-Hawking state through the Euclidean path integral~\cite{HartleHawking1976, GibbonsHawking1977}, extended the diagnosis to finite temperature through the free energy of a static pair in the AdS$_2$ black-hole background, where the subtracted potential again vanishes as the charges are separated.

To put the theory on the lattice, we introduced a covariant discretization scheme for dynamical fermions in curved space in which the spin and gauge connections are packaged together with the kinetic term into a single covariant derivative before discretizing. Because the kinetic and spin-connection terms are separately non-Hermitian and become Hermitian only in combination, this construction is manifestly Hermitian at finite lattice spacing and reproduces the correct continuum Hamiltonian, resolving ambiguities in the existing curved-space lattice fermion literature~\cite{ikeda2025quantumsimulationfermionsads2, ikeda2026geometryinducedchiraltransport}. Using this discretization together with a matrix product state ansatz optimized by DMRG~\cite{White:1992zz, Schollwock:2010zz}, we validated the predicted screened phase from first principles: after numerically subtracting the position-dependent self-energies, the static potential saturates in both the Schwarzschild and global frames, in quantitative agreement with the continuum prediction. As a local check, we extracted the electric flux-tube profile and matched it to the continuum bosonization, finding in both frames a field that changes sign across the insertions and decays exponentially into a screening cloud rather than forming the spatially constant field of a confining string.

Finally, we extended our results beyond the exactly solvable point. Turning on a fermion mass, for which the theory is analytically tractable only in the small-mass expansion~\cite{Alimohammadi:2000fg}, the DMRG simulations showed that the screened phase persists. As a fully nonperturbative, real-time demonstration, we prepared a gauge-invariant meson, an electron-positron pair joined by an electric string, in the global AdS$_2$ vacuum and evolved it with the Time-Dependent Variational Principle (TDVP) algorithm~\cite{Haegeman:2011, Haegeman:2016}. The string breaks as dynamical pairs are produced from the vacuum and the sources are screened, giving a direct real-time picture of Schwinger pair production~\cite{Schwinger:1962tp} and string breaking on a curved background~\cite{BuyensEtAl2014}.

\section*{Acknowledgments}
This work was supported by the Mani L. Bhaumik Institute for Theoretical Physics (S.B. and J.I.) and the National Science Foundation under Grant No.~PHY-2515057 (Z.K.). We thank Per Kraus for numerous insights and helpful discussions, Kazuki Ikeda for useful discussions, and Eric D'Hoker and Dikshant Rathore for comments on the manuscript.

\appendix

\section{Gravitational gauge invariance: continuum and lattice}\label{app:GaugeInvariance} 
In this section, we explain how diffeomorphism invariance is maintained by the fermionic Hamiltonians first on the continuum, and then, in a suitable sense, on the lattice. 

\subsection{Continuum diffeomorphism invariance} \label{app:cont-diff-inv}
Naturally, we expect the continuum Hamiltonian $H_f$ to be diffeomorphism invariant under orientation-preserving coordinate transformations of the kind $r\mapsto R = R(r)$ with Jacobian $J = R'(r)>0$ as they were obtained from the spin connection. Nonetheless, we check this explicitly. The fermions satisfy the canonical commutation relations:
\begin{align}
    \{\chi(r), \chi^\dagger(s)\} = \delta(r-s)
\end{align}
which are maintained under coordinate transformations $r\mapsto R = R(r)$ if and only if:
\begin{align}
    \chi(r)\mapsto J^{-\frac12}\chi(R)
\end{align}
since $\delta(r-s) = J^{-1}\delta(R-S)$. 
Hence, the requirement of diffeomorphism invariance of the Hamiltonian:
\begin{align}
    H_{f} &= -\int dr\;\chi^\dagger\left(if(r)\sigma^x(\partial_r +iA_r)+ i\frac{r}{L^2}\sigma^x  - m\sqrt{f(r)}\sigma^z+\mu\sqrt{f(r)}\right)  \chi
\end{align}
boils down to gauge-covariance of the covariant derivative, i.e.
\begin{align}
    \nabla^{(f)}_r &= f(r)\left(\partial_r + i A_r\right) + \frac12f'(r)
\end{align}
Consider $v = v^r\partial_r$ as a vector field that transforms as $v^R = R'(r)v^r = Jf$. Likewise, $A_R = J^{-1}A_r$ and $\partial_R = J^{-1}\partial_r$. Then we explicitly compute
\begin{align}
    \nabla_R^{(f)}\chi(R)&= \left[v^R\left(\partial_R + i A_R\right) + \frac12\partial_R v^R \right]\chi(R)\nonumber\\&=\left[J(r)f(r)\left(\frac1J\partial_r + i \frac{1}J{}A_r\right) + \frac1{2J}\partial_r\left(J(r)f(r)\right)\right]\left(J^{-\frac12}\chi(r) \right)\nonumber\\
    &= J^{-\frac12}\left[f(\partial_r+iA_r)\chi + \frac12\partial_rf \chi\right] = J^{-\frac12}\nabla_r^{(f)}\chi.
\end{align}
Hence, we obtain:
\begin{align}
    \chi(R)^\dagger  \nabla_R^{(f)}\chi(R)dR = \chi(r)^\dagger  \nabla_r^{(f)}\chi(r)dr
\end{align}
The invariance of the other terms in the Hamiltonian is manifest.

\subsection{Lattice realization of diffeomorphism invariance}
Consider a lattice change of variables that maps the same physical lattice into different variables according to $r_n\mapsto R_n = R(r_n)$. Then the finite difference Jacobian is expressed in terms $\omega_n = \frac{1}{a}(R_{n+\frac12} - R_{n-\frac12})$, which implies that 
\begin{align}
    \chi_n\mapsto\tilde \chi_n =\left(\frac{a}{\omega_n}\right)^{\frac12}\chi_n.
\end{align}
To obtain the transformation of the link objects $f_{n\pm \frac12}$, we consider the covariant derivative operator in the transformed variables, noting that $U_n =\exp{i \int_{n}^{n+1} dr A_r}$ is manifestly coordinate invariant, i.e. $A_r\mapsto A_r$. In the $R$-coordinates, write
\begin{align}
    \tilde \chi_n^\dagger\Delta_{n, m}^{(R)}\tilde\chi_m = \tilde \chi_n^\dagger\frac{F_{n+\frac12}U_n\delta_{m, n+1} - F_{n-\frac12}U_{n-1}^\dagger \delta_{m, n-1}}{2a}\tilde\chi_m \stackrel{\rm!}{=}\chi_n^\dagger\Delta_{n, m}^{(R)}\chi_m 
\end{align}
which can be achieved by the choice: 
\begin{align}
    F_{n\pm\frac{1}{2}} = \frac{\sqrt{\omega_n\omega_{n+1}}}{a}f_{n\pm\frac{1}{2}}\,.
\end{align}
This transformation law of the link objects ensures that the Hamiltonian is gauge invariant. The continuum limit of the above expression is $F(r) = J(r) f(r)$, which is consistent with $v^R = J(r)v^r$ in the continuum case.

\bibliographystyle{JHEP}
\bibliography{refs}
\end{document}